\documentclass[10pt,conference]{IEEEtran}
\usepackage{cite}
\usepackage{amsmath,amssymb,amsfonts}
\usepackage{algorithmic}
\usepackage{graphicx}
\usepackage{textcomp}
\usepackage{xcolor}
\usepackage[hyphens]{url}
\usepackage{fancyhdr}
\usepackage{hyperref}
\usepackage{lipsum}
\usepackage{diagbox}
\usepackage{multirow}
\usepackage{tabularx}
\usepackage{booktabs}
\usepackage[framemethod=tikz]{mdframed}
\usepackage{formatting/macros}
\usepackage{formatting/results}
\usepackage{subfig}
\usepackage{algorithmic}
\usepackage[ruled,vlined]{algorithm2e}
\usepackage{soul}
\usepackage{balance}
\usepackage{xcolor}
\usepackage{hyperref}
\hypersetup{
    colorlinks=true,  
    citecolor=blue,   
    linkcolor=red,    
    urlcolor=teal     
}
\usepackage{makecell}
\usepackage{color, soul}
\usepackage{amsmath, amssymb}
\usepackage{etoolbox}
\renewcommand{\footnoterule}{%
  \kern -3pt
  \hrule width 0.25\textwidth height 0.4pt
  \kern 2.6pt
}

\makeatletter
\def\blfootnote{\gdef\@thefnmark{}\@footnotetext}
\makeatother

\title{Cascade: Exploiting SLO-Aware latency budget for fair and high goodput LLM inference serving}

\def\hpcacameraready{} 

\newcommand\hpcaauthors{%
  \makebox[0.23\textwidth]{Muhammad Adnan$^{\dagger\,\ast}$}%
  \makebox[0.23\textwidth]{Rohan Mahapatra$^{\ddagger}$}%
  \makebox[0.23\textwidth]{Prashant Nair$^{\dagger}$}%
  \makebox[0.23\textwidth]{Daniel Berger$^{\ddagger}$}\\[3pt]
  \makebox[0.23\textwidth]{Pantea Zardoshti$^{\diamond}$}%
  \makebox[0.23\textwidth]{Rodrigo Fonseca$^{\ddagger}$}%
  \makebox[0.23\textwidth]{Esha Choukse$^{\ddagger}$}%
  \\[-0.6em]
}

\newcommand\hpcaaffiliation{%
  \makebox[0.32\textwidth]{$^{\dagger}$The University of British Columbia}%
  \makebox[0.32\textwidth]{$^{\ddagger}$Microsoft Azure Research}%
  \makebox[0.32\textwidth]{$^{\diamond}$NVIDIA}%
}

\author{
  \ifdefined\hpcacameraready
    \IEEEauthorblockN{\hpcaauthors{}}
      \IEEEauthorblockA{
        \hpcaaffiliation{} \\
      }
  \else
    \IEEEauthorblockN{\normalsize{HPCA \hpcayear{} Submission
      \textbf{\#\hpcasubmissionnumber{}}} \\
      \IEEEauthorblockA{
        Confidential Draft \\
        Do NOT Distribute!!
      }
    }
  \fi 
}

\fancypagestyle{camerareadyfirstpage}{%
  \fancyhead{}
  
  \fancyhead[C]{}
  \fancyfoot[C]{}
}

\begin{document}
\maketitle

\blfootnote{$^{\ast}$Work done while interning at Microsoft Azure Research.}
\ifdefined \hpcacameraready 
  \thispagestyle{camerareadyfirstpage}
  \pagestyle{empty}
\else
  \thispagestyle{plain}
  \pagestyle{plain}
\fi

\newcommand{\hpcaheight}{0mm}
\ifdefined\eaopen
\renewcommand{\hpcaheight}{12mm}
\fi


\begin{abstract}
The reasoning and agentic capabilities of LLMs have expanded the range of applications they support, from short interactive exchanges to long, prefix-heavy requests.
LLM serving platforms today define response-latency service-level objectives (SLOs), yet requests under the same SLO differ by orders of magnitude in input length, generation length, execution cost, and reusable KV-cache availability, leaving each request with a different amount of latency headroom.
This headroom must cover not only execution, but also scheduling and preemption delays and the cost of retrieving KV state from deeper memory tiers.
Existing systems manage these decisions independently, despite their combined impact on latency.

We present \cascade, an LLM serving system that coordinates these decisions using a \emph{per-request latency budget}: the difference between a request's SLO and its predicted remaining service time, continuously updated from request characteristics, KV-cache state, and system load.
\cascade uses this budget to prioritize requests and to decide how much non-resident KV state to prefetch from deeper tiers, with the remainder fetched on demand or recomputed given serving system.
On production traces across three LLMs, \cascade improves goodput by \avggoodputimprvllmflatall over \textsc{vLLM} and \avggoodputimprqoserveflatall over a SLO-aware scheduler, and reduces SLO violations by \avgsloviolationredvllmflatall and \avgsloviolationredqoserveflatall, respectively, while remaining fair across heterogeneous requests.
\end{abstract}

\section{Introduction}

Production LLM serving increasingly multiplexes heterogeneous applications~\cite{gemini, code} over
shared model deployments. A single cluster may concurrently serve interactive
chats, coding assistants, tool-using agents, and multi-step reasoning
workloads. While consolidating these services can improve the utilization of
accelerators, it also places requests with widely different execution
demands on the same hardware. Thus, providers need to balance two goals; they must satisfy the latency service level objective (SLO) of each service while keeping shared accelerators efficiently utilized.

Although the SLO is defined at the service level, the work required to serve a
request is inherently request-specific. Requests within the same service can
differ substantially in prompt length, generation length, and model execution
time. Their cost also depends on runtime state. For instance, a request may reuse the Key-Value (KV) state already resident in high-bandwidth memory (HBM), retrieve it from CPU DRAM or
NVMe, or recompute the corresponding prefix. Moreover, output length is unknown
when a request arrives, and reasoning or coding assistants may generate
substantially more work than their inputs suggest. Thus, requests
governed by the same service-level SLO can have very different execution costs
and, therefore, very different urgency.

Model execution is only one component of response latency. A request may
wait in an admission queue, contend with active prefill and decode operations,
and stall while reusable KV state is restored into HBM. Existing systems~\cite{dynamo,gemini} address
these delays through largely independent mechanisms. vLLM~\cite{vllm}, for
example, combines continuous batching and chunked prefill~\cite{orca,sarathi}
with first-come, first-served (FCFS) scheduling. FCFS preserves arrival order,
but allows expensive requests to create severe head-of-line (HOL) blocking.
Shortest-job-first (SJF) scheduling reduces such blocking by advancing shorter
requests, but can systematically deprioritize long-context code and reasoning
workloads. Earliest-deadline-first (EDF)~\cite{edf} scheduling incorporates service
deadlines, but cannot distinguish requests that share a deadline while requiring
substantially different amounts of remaining work.

KV-cache management is typically handled separately from request scheduling.
Prefix-caching systems extend effective KV capacity by retaining reusable state
across requests and spilling it from HBM into deeper memory
tiers~\cite{sglang,lmcache,mooncake}. These systems commonly decide what to
retain, evict, or restore based on cache capacity, reuse probability, and
transfer cost. Such criteria indicate whether a cached object is beneficial in
aggregate, but not whether the request that reuses it can tolerate the resulting
delay. A deep-tier KV-cache hit may therefore avoid expensive prefill
computation while still cause SLO violation because restoring the state
takes longer than the request can afford.
Existing SLO-aware schedulers~\cite{qoserve, sloserve, polyserve} neglect prefix KV data movement overheads and some~\cite{mooncake} reject certain requests to meet SLOs, whereas KV-aware schedulers~\cite{strata} optimize solely for cache migration without providing SLO guarantees.

We make the observation that serving overheads, including queueing, resource contention, preemption, and prefix KV-cache movement, all consume a common resource: the per-request latency budget, defined as the difference between a request's TTFT SLO and its predicted prefill service time. Because this budget varies across requests, the same delay may be tolerable for one request but cause another to violate its SLO. 
Thus, the system should direct unavoidable overhead toward requests with sufficient budget while protecting those with little budget remaining.

\begin{figure*}[t!]
  \centering
  \includegraphics[scale=1]{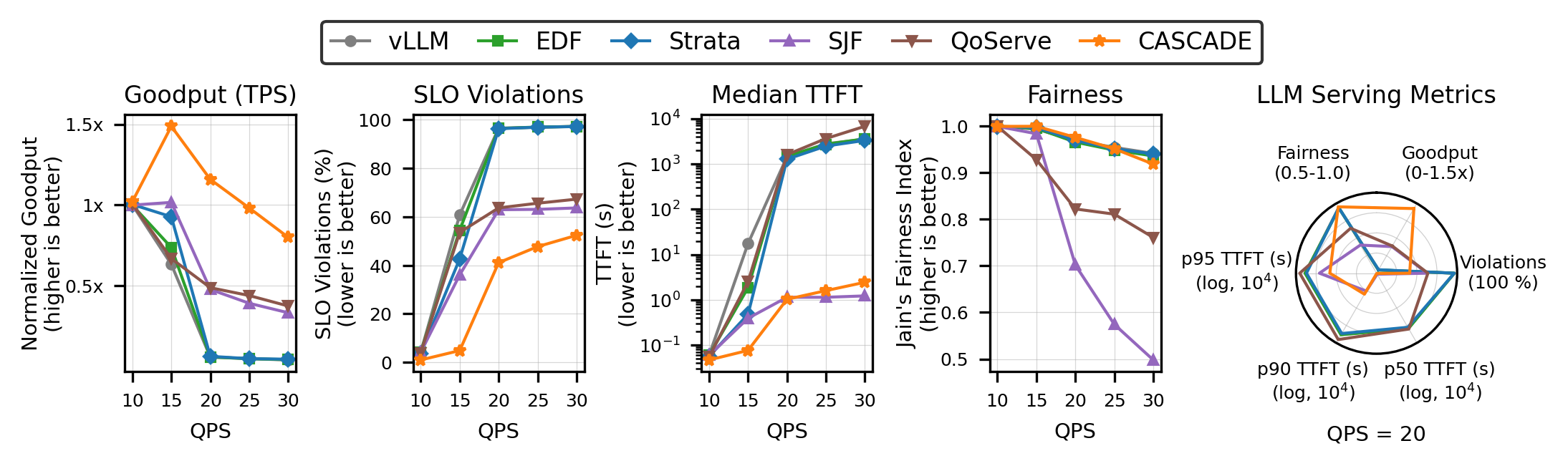}
   \caption{Comparison of state-of-the-art scheduling policies. The graphs plot Goodput, SLO violations, TTFT latency and fairness trade-offs across scheduling policies. We compare
  normalized goodput relative to \textsc{vLLM}(FCFS) with lowest QPS and cross-class fairness, measured using
  Jain's Fairness Index over input-length classes~\cite{fairness_metric}, under a
  \emph{Mixed} production trace~\cite{alibaba_traces} and Qwen-2-72B model. \textsc{Cascade} provides best balance across all LLM serving metrics.}
   \label{fig:intro}
\end{figure*}

\textbf{Our work.}
We present \cascade, an LLM serving system that treats per-request latency budget as a unified abstraction for request scheduling, prefix KV data movement all while complying with SLOs.

\cascade derives the request's latency budget from its SLO target and remaining service time and continuously updates the estimate as the request waits, executes, and encounters changes in system and cache state.
\cascade prioritizes requests based on their remaining
latency budget and request's characteristics under current system load rather than by arrival order, deadline, input size and prefix KV data movement.
Requests with tight remaining latency budget advance ahead of requests that can safely tolerate added delays.
This reduces HOL blocking without assuming
that shorter requests are always more urgent and, unlike size-based scheduling, does not systematically deprioritize long requests.

The same budget also governs prefix KV-cache movement across the memory hierarchy.
When prefix KV state is not resident in HBM, \cascade compares the cost of restoring it against both the request's remaining latency budget and the cost of recomputation.
It prefetches from a deeper tier only if the request can absorb the transfer delay under the current system load; otherwise, it recomputes the prefix.
The budget also determines which KV state to fetch from which tier, since tiers differ in link bandwidth.
Because scheduling and KV movement draw on the same budget, \cascade prevents either from consuming latency budget the other needs.

Relying on any single LLM serving metric, whether goodput, SLO violations, or median latency, can be misleading given the heterogeneity of workloads and requests. These metrics must be considered together with fairness.
Prior work~\cite{vtc} studies fairness across clients; we instead introduce fairness across request lengths, to test whether a scheduling policy treats short and long requests equitably.
Figure~\ref{fig:intro} compares serving metrics and fairness for state-of-the-art scheduling policies under varying load. At high load, SJF and QoServe achieve better goodput and fewer SLO violations, but only by being unfair to long-context requests.

We implement \cascade on top of the vLLM~\cite{vllm} serving engine and extend Vidur~\cite{vidur} to model cluster-scale scheduling and multi-tier KV movement, using profiles collected on NVIDIA GB300 NVL72~\cite{gb200, mnnvl} hardware.
We evaluate \cascade on \emph{production traces} of ChatBot, API, Coder, and Reasoning requests~\cite{alibaba_traces} across Qwen-2-72B, Llama-3-70B, and Llama-3-405B, and on the \emph{Tool\&Agent} trace from Mooncake~\cite{mooncake} to cover agentic workloads.
On average, \cascade improves goodput (SLO-satisfied throughput) by \avggoodputimprvllmflatall over \textsc{vLLM} and \avggoodputimprqoserveflatall over the SOTA SLO-aware QoServe, and reduces SLO violations by \avgsloviolationredvllmflatall and \avgsloviolationredqoserveflatall, respectively, while remaining fair across heterogeneous request classes.

\textbf{Summary.} This paper makes the following contributions:

\begin{itemize}
    \item We identify the \emph{per-request latency budget} as a unifying abstraction for LLM serving: queueing, preemption, execution, and access to reusable KV state in deeper memory tiers all draw on this budget, so they should be optimized jointly rather than independently.

    \item We present \textsc{Cascade}, an LLM serving system that continuously estimates each request's latency budget and uses it to coordinate request scheduling and KV-cache decisions across deeper memory hierarchies.

    \item We introduce fairness across request-length classes as a serving metric, capturing whether a scheduling policy treats short and long requests equitably.

    \item We show that no single metric captures the trade-offs among scheduling policies. By evaluating goodput, SLO violations, latency, and fairness together, we expose behaviors that are hidden when each is considered alone.
\end{itemize}

\section{Background}

\subsection{LLM Inference Serving}
Large Language Model (LLM) inference executes in two asymmetric phases that govern system-level bottlenecks: \textbf{Prefill Phase} is when the model processes the entire input prompt in a single parallel pass to generate the initial Key-Value (KV) cache. This phase is highly compute-intensive, dominated by dense matrix multiplications. \textbf{Decode Phase} is when the model auto-regressively generates output tokens one-by-one, appending each new token's keys and values to the active KV cache. This sequential, token-by-token execution is heavily memory-bandwidth bound.

\paragraph{Service Level Objectives (SLO)}
Service providers define phase-specific Service Level Objectives (SLOs) to maintain user interactivity across distinct execution phases. The prefill phase is governed by \textit{Time to First Token} (TTFT)—the latency between request arrival and first token generation—while the decode phase is bounded by \textit{Time Per Output Token} (TPOT), the latency between successive tokens. In multi-tenant datacenters, these SLOs must be met across heterogeneous workloads with highly varied context and generation lengths, including short-context \emph{API} calls, multi-turn \emph{ChatBot} conversations, long-context \emph{Code} generation, and deep \emph{Reasoning} tasks. Consequently, the primary objective of datacenter-scale LLM serving is to maximize \emph{goodput} -- defined as the volume of request throughput that strictly adheres to these interactive TTFT and TPOT SLO thresholds.

\paragraph{Prefix Caching}
Beyond intra-request reuse, \textit{prefix caching} enables KV cache sharing across distinct requests that share common token sequences. For instance, these could be system prompts, Retrieval-Augmented Generation (RAG) context, or shared code repositories. In multi-turn interactive conversations, prefix caching avoids redundant context computation by storing the concatenated inputs and outputs of previous turns in deeper memory tiers for subsequent requests~\cite{lmcache}. By serving shared prefixes directly from memory rather than recomputing them, this technique shifts the bottleneck of identical context segments from compute-intensive prefill to memory retrieval. This helps to substantially reduce TTFT and increase overall serving throughput.

\subsection{Datacenter-Scale Serving}
Datacenter-scale serving systems deploy multiple model instances to handle diverse user workloads. A centralized \textit{Global Scheduler} routes incoming queries to specific instances based on real-time load balancing and localized HBM prefix cache affinity.

\paragraph{Prefill-Decode Co-location}
Throughout this paper, we focus on co-located LLM serving, where prefill and decode execution share the same serving instances. Although prefill-decode (PD) disaggregation has recently emerged to isolate resource contention across specialized instances~\cite{splitwise, distserve, disaggregatellm}, PD co-location remains an important deployment paradigm in production environments~\cite{aegaeon}. Co-located architectures eliminate complex instance-role management, and avoid the high network transfer overheads of migrating KV caches across physical nodes~\cite{aggregate_serving, tokenscale}. Our findings and policies are generally orthogonal to the choice of co-location and the disaggregation of prefill and decode.

\paragraph{Serving Engine and Instance}
An instance is the minimum hardware unit that serves requests and host one complete copy of the LLM's parameters, potentially spanning multiple GPUs to accomodate model parameters on GPU HBM and serving engines mostly used engines like vLLM~\cite{vllm} and SGLang~\cite{sglang} once an incoming request is pushed into the queue of the instance, queued request is batch and executed efficiently on the same instance utilizing \textit{chunked prefills}~\cite{sarathi}. By partitioning resource-intensive prefill requests into equal-sized chunks, the engine co-schedules prefill chunks alongside active decode steps without inducing execution pauses. This approach eliminates scheduling bubbles, balancing the trade-off between serving latency and system throughput.

\paragraph{Multi-Tier KV Cache Memory Hierarchy}
To satisfy stringent latency SLOs, traditional prefix KV reuse mechanisms~\cite{promptcache, sglang, preble} have primarily relied on local GPU HBM and host CPU DRAM. However, emerging long-context workloads~\cite{kvquant}, such as RAG and multi-step coding agents, rapidly exceed local memory capacity, driving the adoption of global, multi-tier KV memory pools~\cite{superinfer} that extend across NVMe storage and disaggregated network memory~\cite{lmcache, mooncake}. Modern AI cluster architectures facilitate this transition by incorporating ultra-high-bandwidth intra- and inter-node interconnects to accelerate cross-tier state movement. For instance, rack-scale architectures like NVIDIA's GB200/GB300 NVL72~\cite{gb200} feature Multi-Node NVLink (MNNVL)~\cite{mnnvl} across all 18 compute nodes alongside NVLink-C2C for direct host-to-device memory. 

\section{Motivation}

A request's SLO must cover not only model execution but also the queueing,
batching, and KV cache retrieval that serving adds on top of it. The room an SLO
leaves for this overhead is the request's latency budget. Current serving systems
do not track this budget: they let queueing consume it, spend it on KV cache
transfers that exceed it, and report aggregate SLO attainment that hides which
requests run out of it. In this section, we characterize and examine each of these effects in turn.

\subsubsection{Structural Fairness in Multi-Class Request Scheduling}
While policies such as Shortest Job First (SJF) minimize total SLO violations by prioritizing short-context queries, they introduce severe starvation for long-context requests (refer Figure~\ref{fig:prompt_slo_violation}). A single cluster-wide SLO number hides this bias, because the many short requests that succeed outweigh the long-context requests that fail, such as coding agents and multi-turn reasoning. Therefore, a robust scheduler must maintain cross-class fairness, ensuring equitable service quality across all context lengths. 

To quantify cross-class fairness, we employ Jain's Fairness Index ($\text{JFI}$)~\cite{fairness_metric} over request context length bins. Specifically, we partition incoming traffic into six input sequence length percentile ranges: $p_{0}\text{--}p_{50}$, $p_{50}\text{--}p_{75}$, $p_{75}\text{--}p_{90}$, $p_{90}\text{--}p_{95}$, $p_{95}\text{--}p_{99}$, and $p_{99+}$. Let $N$ denote the set of percentile bins ($n = |N|$). For each bin $i \in N$, we define its SLO attainment rate as $A_i = \max(0, 100 - V_i)$, where $V_i$ represents the SLO violation rate of class $i$. The overall fairness index $\mathcal{J}$ is formulated as:
\begin{equation}
\label{eq:jfi}
\mathcal{J} = \frac{\left( \sum_{i=1}^{n} A_i \right)^2}{n \sum_{i=1}^{n} A_i^2}
\end{equation}
The index spans $\mathcal{J} \in [\frac{1}{n}, 1.0]$, where $\mathcal{J} = 1.0$ represents ideal fairness (identical SLO attainment across all context length bins), whereas $\mathcal{J} = \frac{1}{n}$ indicates maximum unfairness, where a single bin monopolizes all successful completions while others suffer total non-compliance.

\finding{
\takeaway{1}{\textbf{A high overall SLO attainment can still hide starved request
classes.} A scheduler can meet its SLO for most requests while failing nearly all
long-context requests, because short requests dominate the count. Measuring SLO
attainment separately for each prompt-length range exposes this gap, which a single
cluster-wide number conceals.}}

\subsubsection{The Deep-Tier KV Fetch Dilemma}
\label{sec:motivation_slack}

\begin{figure}[t]
  \centering
  \includegraphics[scale=0.49]{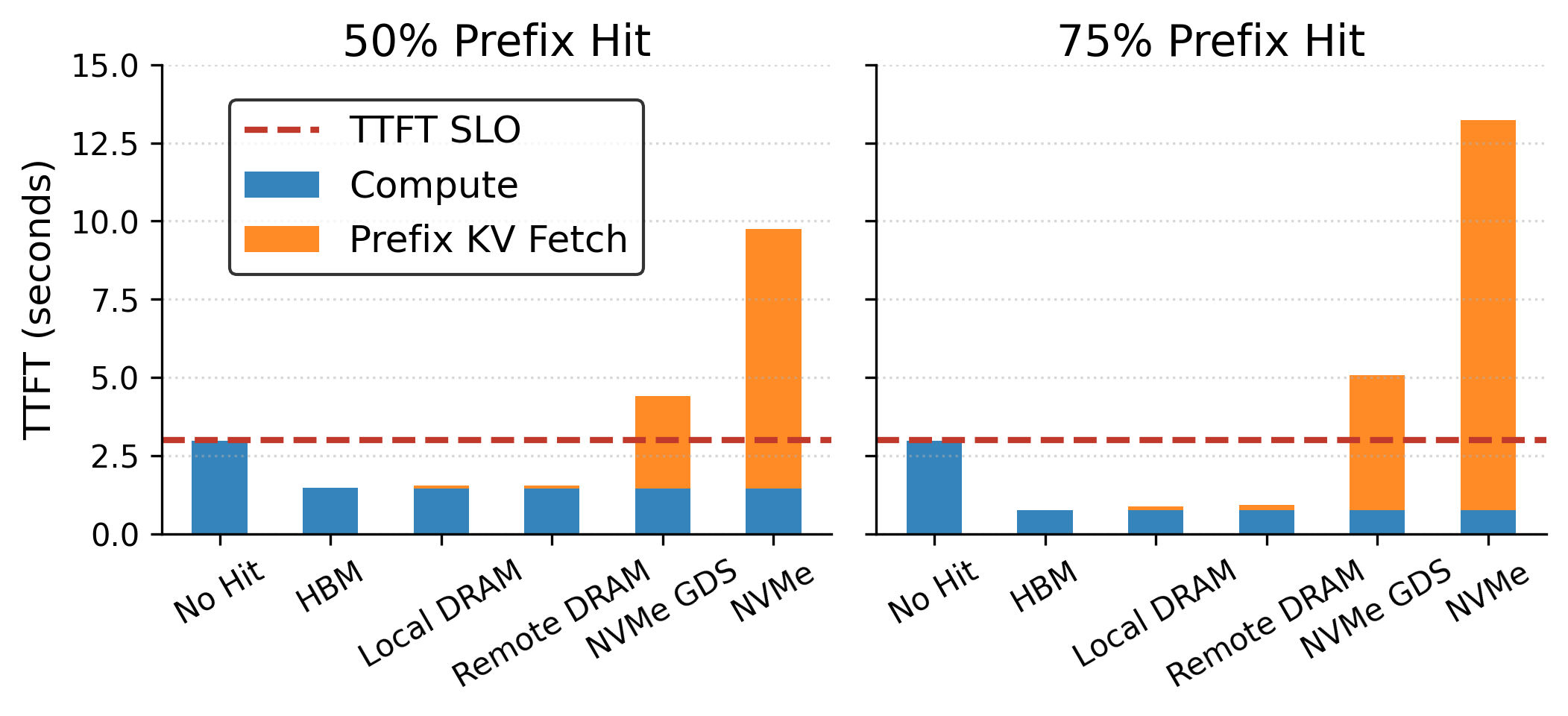}
\caption{Impact of deep-tier KV retrieval on TTFT for a $64\text{k}$ context synthetic request (Qwen-2-72B) on GB300 NVL72. Fetching prefix state from NVMe reduces prefill compute but introduces high transfer latency, turning naive deep-tier hits into TTFT SLO violations.}
\label{fig:kv_fetch_sensitivity}
\end{figure}

While extending prefix KV caching to deep memory tiers (e.g., NVMe drives and
disaggregated storage) expands cache capacity for long-context workloads, it
introduces a fundamental tension between prefill compute savings and state
restoration latency. Figure~\ref{fig:kv_fetch_sensitivity} shows this tension for a
64k-token prefix on Qwen-2-72B. Reusing the prefix from a high-speed tier (HBM or
DRAM) reduces GPU prefill compute by up to 75\% with negligible transfer cost, which
leaves most of the TTFT SLO unspent. Retrieving the same prefix from NVMe, however,
adds transfer delay along the NVMe to CPU DRAM to GPU path that exceeds the TTFT SLO,
even with LMCache batched prefetching~\cite{lmcache}, an effect also reported in prior
work~\cite{tutti}. A runtime that admits every deep-tier hit
therefore converts some cache hits into SLO violations.

\finding{
\takeaway{2}{\textbf{Deep-tier prefix caching requires budget-aware restore admission.} To safely harness disaggregated KV memory pools, serving schedulers must dynamically assess whether a request's remaining latency budget can absorb transfer delays, falling back to GPU prefill recomputation whenever transfer overhead threatens SLO compliance.}}

\subsubsection{Reclaiming Intrinsic Request Latency Budget}
\label{sec:motivation_slack_concept}

\begin{figure}[t]
  \centering
  \includegraphics[width=1\columnwidth]{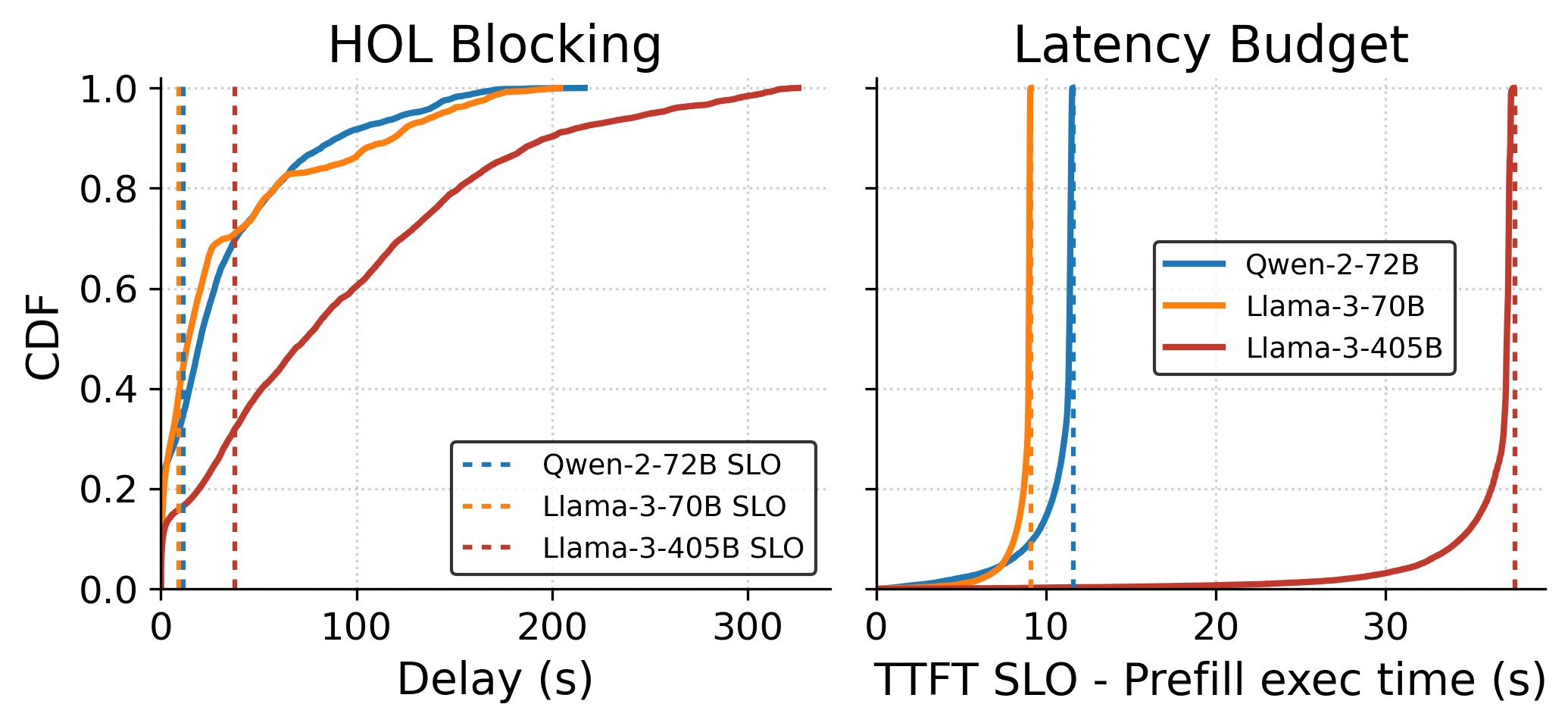}
\caption{Impact of Head-of-Line (HOL) blocking under \textsc{vLLM} scheduling for \emph{Tool\&Agent} trace. \textsc{vLLM}(FCFS) queuing introduces severe HOL blocking delays for arriving requests, rapidly consuming their latency headroom. In contrast, evaluating requests without HOL delay reveals substantial intrinsic budget that can be exploited for SLO-compliant KV prefetching and request scheduling.}
\label{fig:delay_vs_slack}
\vspace{-0.2in}
\end{figure}

Conventional LLM serving schedulers rely on First-Come-First-Served (FCFS) or size-agnostic queuing, subjecting arriving requests to severe Head-of-Line (HOL) blocking delays. As shown in Figure~\ref{fig:delay_vs_slack} (left), FCFS queuing introduces massive delays—frequently exceeding tens of seconds—that rapidly exhaust a request's Service Level Objective (SLO) headroom before GPU execution even commences. However, evaluating these requests in isolation reveals a crucial opportunity: requests inherently possess substantial \emph{latency budget}, defined as the margin between a request's target Time-To-First-Token ($\text{TTFT}$) SLO and its intrinsic GPU prefill execution time ($\text{Budget} = \text{TTFT}_{\text{SLO}} - T_{\text{prefill}}$). As shown in Figure~\ref{fig:delay_vs_slack} (right), unblocked requests across Qwen-2.5-72B, Llama-3-70B, and Llama-3-405B exhibit intrinsic latency budget ranging from $2\text{~seconds}$ to over $20\text{~seconds}$. Rather than allowing HOL queuing to squander this budget, an efficient serving runtime should explicitly model and exploit per-request latency budget -- reordering execution queues and orchestrating deep-tier KV state transfers without violating downstream SLO contracts.

\finding{
\takeaway{3}{\textbf{SLO violations stem from HOL blocking, not a lack of intrinsic budget.} Requests across model scales possess multi-second latency budget ($\text{TTFT}_{\text{SLO}} - T_{\text{prefill}}$) that legacy FCFS schedulers waste in queues; available budget-aware scheduling reclaims this headroom to coordinate request dispatching and deep-tier KV cache management.}}
\section{Cascade: KV and SLO-aware LLM Serving}

State-of-the-art LLM serving schedulers rely on rigid, static policies that fail to adapt to dynamic workload fluctuations and system overload. When cluster utilization spikes, legacy schedulers either selectively favor short requests at the expense of long-context workloads or induce severe Head-of-Line (HOL) blocking—driving up queuing delays and triggering widespread Time-To-First-Token (TTFT) SLO violations. Simultaneously, naive prefix KV cache retrieval introduces exposed transfer overheads that stall prefill execution and delay co-located decode batches, degrading Time-Per-Output-Token (TPOT). Under GPU HBM exhaustion, standard tail-preemption heuristics further compound latency by indiscriminately evicting recently scheduled requests. 

Resolving these intertwined bottlenecks requires a request-aware serving system that holistically coordinates request dispatch, state movement, and context preemption. Based on these observations, we introduce \cascade, KV and SLO-aware LLM serving system centered around per-request \emph{latency budgets}. \cascade defines a request's latency budget as the difference between its overall SLO target and its intrinsic execution time, safely co-allocating this budget across queuing delays, prefix KV transfers, and preemption decisions to preserve strict Service Level Agreement (SLA) compliance while maximizing cluster goodput.

\begin{figure}[t]
  \centering
  \includegraphics[width=1\columnwidth]{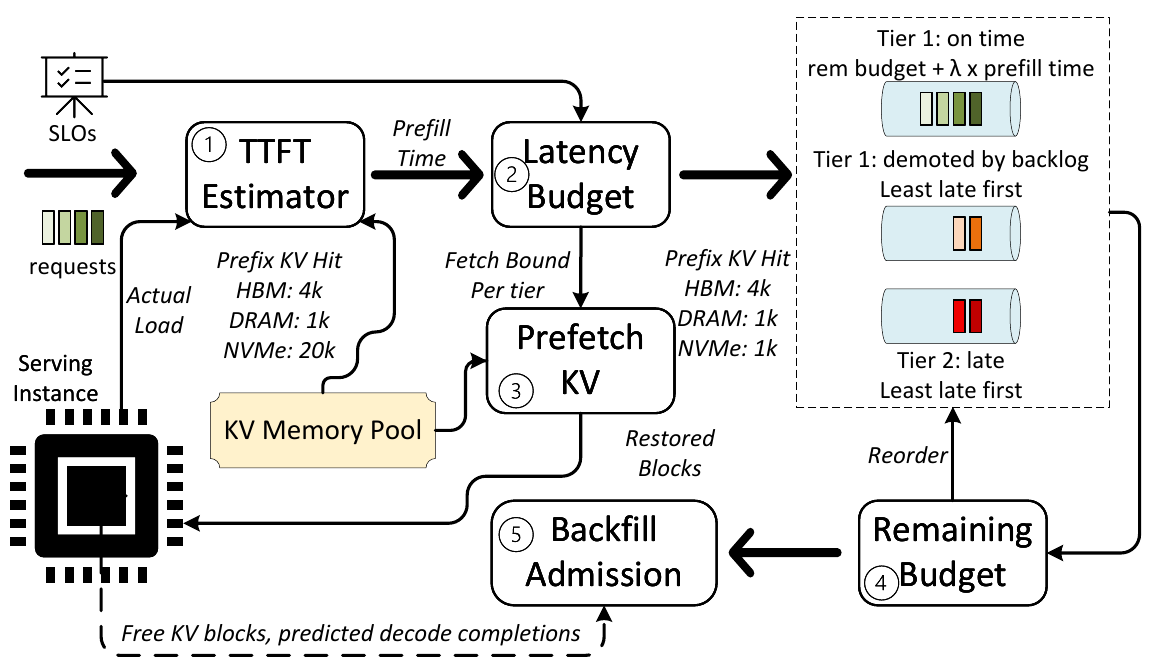}
   \caption{Overview of Cascade.}
   \label{fig:cascade}
\end{figure}

Figure~\ref{fig:cascade} shows the architecture of \textsc{Cascade}. When a request arrives at a serving instance, the \encirclelight{1}~\emph{TTFT Estimator} predicts its prefill time from the instance's current decode load and the prefix KV hit in each memory tier. The \encirclelight{2}~\emph{Latency Budget} is the margin between the request's class TTFT target and this prediction. The budget also bounds how much prefix state each tier can supply in time, and \encirclelight{3}~\emph{Prefetch KV} starts restoring the cached prefix within that bound as soon as the request is queued, so the transfer overlaps with queueing. At every chunked iteration, \encirclelight{4}~\emph{Remaining Budget} subtracts the time each request has waited, measures queue pressure, and reorders the queue. Requests that can still meet their target form Tier-1 and are ordered by remaining budget plus a pressure-weighted prefill time, which moves short requests ahead as the backlog grows. A Tier-1 request that the work ahead of it would make late is demoted behind the others until the backlog drains. Requests that have already missed their target move to Tier-2 and are served after Tier-1, least late first, and are never dropped. Finally, \encirclelight{5}~\emph{Backfill Admission} admits requests in this order. When the request at the head cannot get KV blocks, it is set aside with its blocks reserved, and later requests that fit are admitted in its place. Below, we formalize the serving problem and then detail each component of \cascade with a toy example.

\subsection{Problem Formulation}
\label{sec:problem_formulation}

A cluster serves requests from a set of classes $\mathcal{C}$ (interactive chat, agentic tool use, code generation, reasoning). Class $c$ has a time-to-first-token target $S^{\mathrm{TTFT}}_{c}$ and a time-per-output-token target $S^{\mathrm{TPOT}}_{c}$; for a request $r$ of class $c(r)$ we write $S^{\mathrm{TTFT}}_{r}$ and $S^{\mathrm{TPOT}}_{r}$. Targets are set per class~\cite{mooncake} because prompt and output lengths vary too much within a class for per-request targets to be enforced online. \cascade{} also supports request-length-based targets (Section~\ref{sec:percentile_slo}).

Prefix KV state is stored in blocks of $z$ tokens and $m$ bytes across three tiers: GPU HBM ($k=0$, where requests execute), host DRAM ($k=1$) and NVMe ($k=2$). HBM holds $N^{\mathrm{HBM}}$ blocks. Tier $k \in \{1,2\}$ reaches HBM over a path $P(k)$ of links, and link $\ell$ has BW $W_\ell$.

\cascade{} chooses a scheduling policy $\pi$ (admission order and batch) and a retrieval policy $\psi$ (which cached blocks to restore) that maximize the number of requests meeting both targets:
\begin{equation}
\label{eq:opt_goal_compact}
\begin{aligned}
\max_{\pi,\psi}\ & \sum_{r \in \mathcal{R}} \mathbb{I}\bigl(\mathrm{TTFT}_r \le S^{\mathrm{TTFT}}_{r} \land \mathrm{TPOT}_r \le S^{\mathrm{TPOT}}_{r}\bigr) \\
\text{s.t.}\ & n^{\mathrm{HBM}}(t) \le N^{\mathrm{HBM}}, \quad u_\ell(t) \le W_\ell \quad \forall \ell, t,
\end{aligned}
\end{equation}
where $n^{\mathrm{HBM}}(t)$ is the number of HBM blocks in use and $u_\ell(t)$ the transfer rate on link $\ell$ at time $t$.

\subsection{TTFT Latency Estimator and Latency Budget}
\label{sec:latency_estimator}

When $r$ arrives at time $a_r$, \cascade{} predicts how long its prefill will take and sets that time aside from its TTFT target. A predictor trained offline on execution profiles (as in Vidur~\cite{vidur}) takes the prompt length, the cached prefix found in each tier, and the number of running decodes. It returns $\hat T_r$, the prefill time of the uncached tokens at the current decode load, and $\hat T^{\max}_r$, the same time when the batch holds the maximum of $n_{\max}-1$ decodes. The estimate, the budget, and the remaining budget at time $t$ are
\begin{align}
L_r &= \tfrac{\gamma}{2}\bigl(\hat T_r + \hat T^{\max}_r\bigr), \qquad \gamma \ge 1, \label{eq:est_latency}\\
B_r &= S^{\mathrm{TTFT}}_{r} - L_r, \label{eq:budget_eq}\\
B_r(t) &= B_r - (t - a_r). \label{eq:rem_budget_eq}
\end{align}
$B_r(t)$ is the time $r$ can still spend queued or restoring prefix state and meet its target. The guardband $\gamma$ is fixed offline per model and held constant across traces and load levels. The budget bounds TTFT only. TPOT is protected by chunked prefill, which caps the tokens an iteration processes at $Z$, so admitting a long prompt does not stall the decodes that share its batch.

\paragraph{Misprediction.}
An error in $L_r$ shifts $B_r(t)$ by the same amount, and the queueing term comes from the clock and is exact. Two requests swap order only when their errors differ by more than the gap between their budgets, so a bias shared across requests largely cancels. Under-prediction is the costly direction because it defers a request that has no time to spare, and both $\gamma$ and the average with $\hat T^{\max}_r$ in Eq.~\ref{eq:est_latency} push errors toward over-prediction. 

\subsection{Budget-Feasible Prefix Restoration}
\label{sec:remaining_budget_kv_fetch}

Restoring a cached prefix saves prefill compute but takes transfer time, and that time comes out of the same budget. Let $\delta_\ell$ be the fixed per-transfer overhead of link $\ell$, and let $W_\ell$ here be the bandwidth of $\ell$ that earlier transfers have not already claimed. Restoring $x$ bytes from tier $k$ takes
\begin{equation}
\label{eq:restore_time}
T_k(x) = \sum_{\ell \in P(k)} \delta_\ell + x \sum_{\ell \in P(k)} \frac{1}{W_\ell},
\end{equation}
which charges every link on the path and is therefore an upper bound. A restore is feasible for $r$ if it fits in the budget, so the most that tier $k$ can supply is
\begin{equation}
\label{eq:max_restorable}
M_{r,k} = \frac{\bigl[B_r - \sum_{\ell \in P(k)} \delta_\ell\bigr]^+}{\sum_{\ell \in P(k)} 1/W_\ell}.
\end{equation}
\cascade{} computes $M_{r,k}$ when $r$ is enqueued and starts restoring its cached prefix at once, taking at most $M_{r,k}$ bytes from tier $k$. The transfer runs while $r$ waits, so queueing hides it, and the batch waits only for the part that is still in flight when $r$ is admitted. Cached blocks beyond the bound are decided at admission: \cascade{} cuts the prefix at the point that minimizes fetch time before the cut plus recompute time after it. Following LMCache~\cite{lmcache}, restores run on a separate non-blocking CUDA stream, and a CUDA event synchronizes each chunk before the kernel that consumes it.

\subsection{Overload-Aware Scheduling}
\label{sec:cd}

Consider one replica serving tool-calling agents whose arrival rate alternates between 2.5 and 7.5 requests per second every ten minutes. Most requests are short and a few carry tens of thousands of tokens of context. The replica keeps up in the quiet phase. In the busy phase the queue grows past what it can drain within a TTFT target.
Each iteration splits the waiting queue $\mathcal{Q}$ by remaining budget. Tier-1, $\mathcal{Q}_1 = \{r : B_r(t) \ge -\epsilon\}$ with $\epsilon = 0.1$\,s, holds the requests that can still meet their target. Tier-2, $\mathcal{Q}_2 = \mathcal{Q} \setminus \mathcal{Q}_1$, holds those that have missed it. The base policy admits Tier-1 in ascending $B_r(t)$, then Tier-2 least late first, and stops at the first request whose KV blocks do not fit. Because $B_r$ already subtracts the request's own prefill time, a long request with a loose target keeps a positive budget and is not starved as it would be under shortest-job-first. Under overload this order fails in three ways. \cascade{} adds one rule for each (Sections~\ref{sec:cd_order} to \ref{sec:cd_backfill}), and the first two share one signal, the time the queue needs to drain relative to the TTFT target.

\subsubsection{Cost-Aware Order}
\label{sec:cd_order}

The request with the least budget is usually the longest one, so under overload the base order spends each iteration on the requests that are hardest to finish while short ones expire. On Llama-3-405B, a policy that favors short requests meets 83.4\% of targets on prompts under 1K tokens against 62.8\% for the base order, and gives up the 16K to 64K bucket (16.3\% against 34.2\%). \cascade{} orders Tier-1 by
\begin{align}
\mathrm{key}_r &= B_r(t) + \lambda\,\hat T_r, \qquad \lambda = \min(\lambda_{\max}, \rho), \label{eq:cd_key}\\
\rho &= \frac{1}{\sigma \bar S}\sum_{u \in \mathcal{Q}} \hat T_u, \qquad \sigma = \max\Bigl(\sigma_{\min},\, 1 - \frac{n_{\mathrm{dec}}}{Z}\Bigr). \label{eq:cd_pressure}
\end{align}
Here $\bar S$ is the mean TTFT target in $\mathcal{Q}$, $n_{\mathrm{dec}}$ the number of running decodes, and $\sigma$ the share of an iteration's $Z$ tokens left for prefill ($\sigma_{\min} = 0.25$). The pressure $\rho$ is the queue's drain time in units of the target. On an idle replica $\rho \approx 0$ and the order is the base order. At $\lambda = 1$ it is deadline order, and above 1 it favors short requests; we cap $\lambda$ at $\lambda_{\max} = 2$.

\finding{Example. A flat 9\,s target and a backlog of 10 long requests (6\,s each) and 60 short ones (0.1\,s each). The base order meets one target. Here $\rho = 66/9 > 7$, so $\lambda = 2$: the long requests get key $3 + 12 = 15$, the short ones $8.9 + 0.2 = 9.1$, and all 60 short requests finish within 6\,s.}

\subsubsection{Backlog-Aware Budget}
\label{sec:cd_backlog}

$B_r(t)$ counts only the request's own prefill, not the work queued ahead of it. In the base policy, 38\% (Llama-3-405B) and 54\% (Qwen-2-72B) of TTFT misses had a positive budget at arrival. \cascade{} walks Tier-1 in key order and charges each request the work kept ahead of it:
\begin{equation}
\label{eq:cd_backlog}
D_r = \frac{1}{\sigma}\sum_{u \prec r,\ u\ \text{kept}} \hat T_u,
\qquad
r\ \text{kept} \iff B_r(t) - D_r \ge -\epsilon.
\end{equation}
A request that fails the test is demoted behind the kept ones, least late first. It stays in Tier-1 and the test is repeated every iteration, so it moves back as soon as the backlog ahead of it drains.

\finding{Example. A request with 1\,s of prefill and a 9\,s target arrives with 8\,s of latency budget while 20\,s of work is kept ahead of it. The base order ranks it as feasible for 8\,s. Here $B_r(t) - D_r = 8 - 20 = -12$\,s, so it is demoted until the queue drains.}

\subsubsection{Backfill}
\label{sec:cd_backfill}

Stopping at a blocked head makes every request behind it wait for memory it does not need. On the two overloaded models, 12\% (Llama-405B) and 59\% (Qwen-72B) of TTFT misses waited behind a blocked head. Let $W$ be the first request that does not fit, $N_r$ the HBM blocks $r$ needs at admission, and $F$ the free blocks. A running decode $u$ has $\hat R_u$ predicted output tokens left (the 75th percentile for its class) and frees $n_u$ blocks when it finishes. With mean iteration time $\bar\tau$, the blocks for $W$ are expected at
\begin{equation}
\label{eq:cd_tfree}
t^{\mathrm{free}}_W = \min\Bigl\{\hat R_v \bar\tau : \sum_{u:\, \hat R_u \le \hat R_v} n_u \ge N_W - F\Bigr\},
\end{equation}
and $n^{\mathrm{free}}_W$ blocks are freed by then. \cascade{} sets $W$ aside, reserves its blocks, and keeps admitting. A later request $r$ is admitted only if it fits now and its footprint at $t^{\mathrm{free}}_W$ leaves the reservation intact:
\begin{align}
\Phi &= F + n^{\mathrm{free}}_W - N_W, \label{eq:cd_reserve}\\
N_r \le F \ &\land\ N_r + \Bigl\lceil \tfrac{1}{z}\min\bigl(\hat R_r,\ t^{\mathrm{free}}_W/\bar\tau\bigr) \Bigr\rceil \le \Phi. \label{eq:cd_backfill}
\end{align}
Each admission subtracts its footprint from $\Phi$, and at most $d_{\max} = 16$ blocked requests are set aside per iteration.

\finding{Example. $W$ needs 1{,}900 blocks, 600 are free, and running decodes release 1{,}500 within 3\,s. Then $\Phi = 600 + 1500 - 1900 = 200$: two short requests of about 70 blocks each start now, and $W$ still gets its blocks at 3\,s.}

\begin{figure}[t]
  \centering
  \includegraphics[width=1\columnwidth]{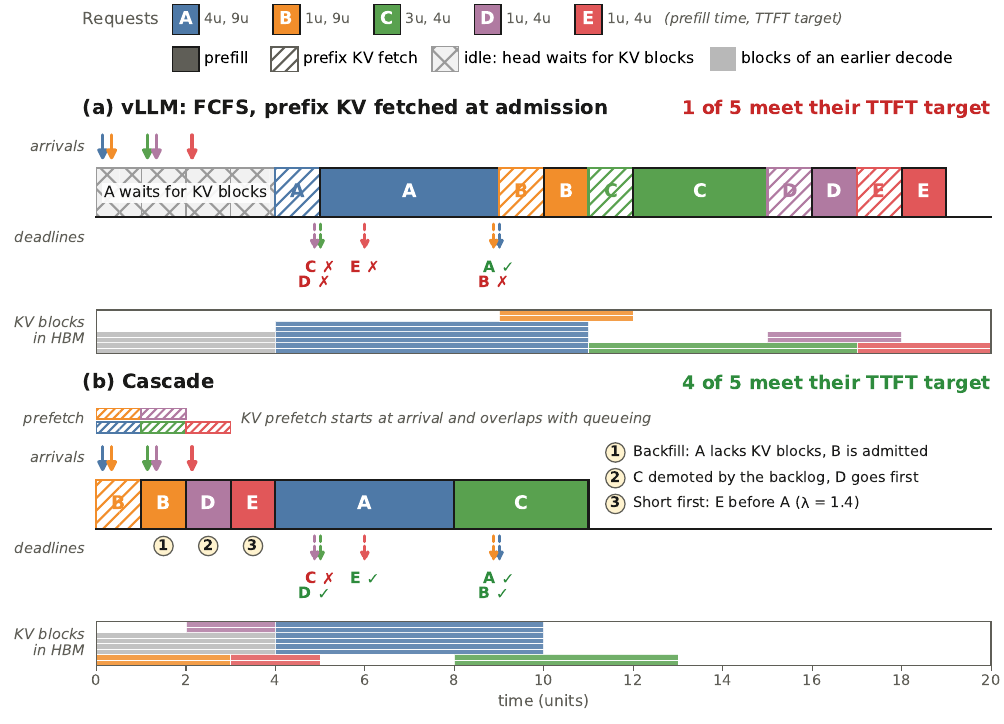}
  \vspace{-0.2in}
   \caption{An illustration of how \cascade perform KV and SLO-aware scheduling.}
   \label{fig:toy_example}
   \vspace{-0.2in}
\end{figure}

\paragraph{Preemption for a blocked request.}
Backfill waits for decodes to finish. A blocked $W$ can instead preempt running requests, which discards their progress. A running request $v$ that has generated $g_v$ tokens has budget

\begin{equation}
\label{eq:cd_run_budget}
\tilde{B}_v =
\begin{cases}
\begin{aligned}
  &a_v + \max\bigl(\mathrm{TTFT}_v, S^{\mathrm{TTFT}}_{v}\bigr) \\
  &\quad + (g_v{+}1)\,S^{\mathrm{TPOT}}_{v} - t,
\end{aligned} & \text{decoding,}\\[1.2ex]
a_v + S^{\mathrm{TTFT}}_{v} - t, & \text{prefilling.}
\end{cases}
\end{equation}

and a resume cost $C_v$, the predicted time to recompute what it has processed. $W$ may preempt only if it can still meet its target but would miss it by waiting:
\begin{equation}
\label{eq:cd_need}
t + \hat T_W \le a_W + S^{\mathrm{TTFT}}_{W} < t + t^{\mathrm{free}}_W + \hat T_W .
\end{equation}
With $B^{\ast}_W = a_W + S^{\mathrm{TTFT}}_{W} - t - \hat T_W$, the eligible victims are
\begin{equation}
\label{eq:cd_victims}
\mathcal{V}_W = \bigl\{v \in \mathcal{R}^{\mathrm{run}} \setminus \mathcal{P} :
\tilde B_v - \mu > B^{\ast}_W,\ C_v + \hat T_W \le \tilde B_v\bigr\},
\end{equation}
with margin $\mu = 0.3$\,s. The protected set $\mathcal{P}$ holds requests that have been preempted twice, are about to finish, or have just resumed. Victims are taken in increasing $C_v/n_v$ until they free $N_W - F$ blocks, all or nothing, for at most one request per iteration. Without a further check this restarts 18\% (Llama-3-405B) and 20\% (Qwen-2-72B) of requests, often recomputing more tokens than the preemption wins. A gate therefore requires
\begin{equation}
\label{eq:cd_gate}
p_W + \hat R_W \ge \kappa \sum_{v \in \mathcal{V}_W} x_v
\end{equation}
and fewer than $K$ preemptions in the last $T_{\mathrm{win}}$, where $p_W$ is the prompt length of $W$ and $x_v$ the tokens $v$ has processed ($\kappa = 1$, $K = 20$, $T_{\mathrm{win}} = 10$\,s). 
Algorithm~\ref{alg:cascade} puts the pieces together.

\begin{algorithm}[t]
\caption{\cascade{} KV and SLO-aware scheduling}
\label{alg:cascade}
\footnotesize
\SetKwFunction{FAdmit}{Admit}
\SetKwProg{Fn}{Proc}{:}{}
\KwIn{waiting queue $\mathcal{Q}$, running set $\mathcal{R}^{\mathrm{run}}$, free blocks $F$}
\KwOut{batch $\mathcal{B}$}
\BlankLine
\tcc{On arrival of $r$}
$B_r \gets S^{\mathrm{TTFT}}_{r} - L_r$ \tcp*{Eqs.~\ref{eq:est_latency}, \ref{eq:budget_eq}}
start restoring $\le M_{r,k}$ bytes from each tier $k$ \tcp*{Eq.~\ref{eq:max_restorable}}
$\mathcal{Q}.\mathrm{push}(r)$\;
\BlankLine
\tcc{Each iteration. 1: running requests}
add each $u \in \mathcal{R}^{\mathrm{run}}$ to $\mathcal{B}$; if HBM is full, preemption\;
\tcc{2: tiers}
$\mathcal{Q}_1 \gets \{r \in \mathcal{Q} : B_r(t) \ge -\epsilon\}$;\ \ $\mathcal{Q}_2 \gets \mathcal{Q} \setminus \mathcal{Q}_1$ \tcp*{Eq.~\ref{eq:rem_budget_eq}}
\tcc{3: order Tier-1 under pressure}
compute $\sigma$, $\rho$, $\lambda$;\ \ sort $\mathcal{Q}_1$ by $B_r(t) + \lambda \hat T_r$ \tcp*{Eqs.~\ref{eq:cd_key}, \ref{eq:cd_pressure}}
$A \gets 0$;\ \ $\mathcal{Q}_1^{+} \gets [\,]$;\ \ $\mathcal{Q}_1^{-} \gets [\,]$\;
\ForEach{$r \in \mathcal{Q}_1$ in order}{
    \lIf{$B_r(t) - A/\sigma \ge -\epsilon$}{append $r$ to $\mathcal{Q}_1^{+}$;\ \ $A \gets A + \hat T_r$}
    \lElse{append $r$ to $\mathcal{Q}_1^{-}$ \tcp*[f]{Eq.~\ref{eq:cd_backlog}}}
}
$\mathcal{O} \gets \mathcal{Q}_1^{+}$, then $\mathcal{Q}_1^{-}$ and $\mathcal{Q}_2$, each least late first\;
\tcc{4: admit with backfill}
$\Phi \gets \bot$;\ \ $s \gets 0$\;
\ForEach{$r \in \mathcal{O}$ \textbf{while} tokens and batch slots remain}{
    \uIf{$N_r \le F$ \textbf{and} ($\Phi = \bot$ \textbf{or} $\mathrm{footprint}(r) \le \Phi$)}{
        \FAdmit{$r$} \tcp*{Eq.~\ref{eq:cd_backfill}}
        \lIf{$\Phi \ne \bot$}{$\Phi \gets \Phi - \mathrm{footprint}(r)$}
    }
    \uElseIf(\tcp*[f]{option}){Eqs.~\ref{eq:cd_need}, \ref{eq:cd_gate} hold and $\mathcal{V}_r$ frees $N_r - F$}{
        preempt and requeue each $v \in \mathcal{V}_r$;\ \ \FAdmit{$r$} \tcp*{Eq.~\ref{eq:cd_victims}}
    }
    \uElseIf{$s < d_{\max}$}{
        \lIf{$\Phi = \bot$}{$\Phi \gets F + n^{\mathrm{free}}_r - N_r$ \tcp*[f]{Eqs.~\ref{eq:cd_tfree}, \ref{eq:cd_reserve}}}
        $s \gets s + 1$;\ \ set $r$ aside\;
    }
    \lElse{\textbf{break}}
}
\BlankLine
\Fn{\FAdmit{$r$}}{
    for cached blocks beyond $M_{r,k}$: fetch or recompute, whichever is faster\;
    $\mathcal{B} \gets \mathcal{B} \cup \{r\}$;\ \ $F \gets F - N_r$\;
}
\end{algorithm}

\subsection{\cascade: Toy Example}
\label{sec:example}

Figure~\ref{fig:toy_example} runs five requests on one replica, in abstract time units. HBM holds 8 KV blocks, and an earlier decode holds 4 of them until $t=4$. Request A is long (4 units of prefill, 6 blocks) with a 9-unit TTFT target. B is short with the same target. C, D and E have a 4-unit target; C needs 3 units of prefill and D and E need 1. Each request has a cached prefix that takes 1 unit to restore.

\paragraph{vLLM.}
FCFS admits in arrival order and fetches the prefix at admission. A arrives first and does not fit, so the replica idles for 4 units while four requests wait behind it. A then pays 1 unit of fetch and finishes at $t=9$, exactly on its target. Every later request pays its own fetch and misses. One of five targets is met, and the last prefill ends at $t=19$.

\paragraph{\cascade{}.}
All five prefetches start at arrival, so only B, admitted at $t=0$, waits for its fetch. \encirclelight{1}~At $t=0$, A has the lowest key but only 4 blocks are free. \cascade{} sets A aside and reserves its blocks: $\Phi = 4 + 4 - 6 = 2$ (Eq.~\ref{eq:cd_reserve}), which admits B. \encirclelight{2}~At $t=2$ the queue holds A, C, D and E, and the pressure is $\rho = 1.7$. C has no budget left and 2 units of work are kept ahead of it, so it is demoted (Eq.~\ref{eq:cd_backlog}) and D goes first. Serving C first, as least-budget order would, makes D, E and A all miss. \encirclelight{3}~At $t=3$, E goes before A because its key is lower (Eq.~\ref{eq:cd_key}). At $t=4$ the earlier decode ends, A gets the 6 blocks reserved for it, and it finishes at $t=8$. C runs last and misses. Four of five targets are met, and the last prefill ends at $t=11$.

\begin{scriptsize}
\begin{table}[t]
\centering
\caption{Traces used in evaluation}
\resizebox{1\columnwidth}{!}{
\begin{tabular}{ l | c | c c c | c c c | c c }
\toprule
\multirow{2}{*}{\textbf{Trace}} & \textbf{No. of} &  \multicolumn{3}{c|}{\textbf{Input Sequence Length}} & \multicolumn{3}{c|}{\textbf{Output Sequence Length}} & \multicolumn{2}{c}{\textbf{Multi-turn}} \\
& \textbf{Requests} & \textbf{p50} &  \textbf{p99} & \textbf{Max} & \textbf{p50} & \textbf{p99} & \textbf{Max} & \textbf{\% Reqs} & \textbf{Max Turns} \\
\midrule
ChatBot   & 43,058 &  1,046 &  14,364 & 89,286 & 375   &  1,641  & 8,000 & 46.35 & 37 \\
API & 172,800 &  574   &  6,294  & 66,446 & 39    &  1,006  & 8,192 & 0.00 & 1 \\
Coder & 43,011 &  4,540 &  14,406 & 25,777 & 469   &  5,842  & 35,031 & 38.61 & 54 \\
Reasoning & 10,812 &  3,680 &  24,972 & 51,622 & 1,665 &  34,686 & 75,990 & 11.10 & 16 \\
\midrule
Mixed & 269,681 &  886 &  13,792 & 89,286 & 86 &  4,812 & 75,990 & 14.00 & 54 \\
ChatBot Heavy   & 54,390 &  1,016 &  14,188 & 89,286 & 328   &  2,178  & 65,124 & 38.35 & 44 \\
API Heavy & 73,965 &  627   &  11,183 & 66,446 & 47    &  1,671  & 65,124 & 2.59 & 44 \\
Coder Heavy & 54,345 &  2,883 &  14,311 & 59,159 & 348   &  5,711  & 65,124 & 32.54 & 54 \\
Reasoning Heavy & 23,756 &  1,298 &  22,301 & 59,159 & 425 &  22,248 & 75,990 & 12.86 & 44 \\
Uniform & 41,923 & 1,430 & 19,081 & 82,770 & 390 &  14,630 & 75,990 & 23.75 & 48 \\
\midrule
Tool\&Agent & 23,608 & 6,858 & 62,173 & 126,707 & 30 &  898 & 2,000 & 30.6 & 43 \\
\bottomrule
\end{tabular}}
\label{tab:traces}
\end{table}
\end{scriptsize}

\subsection{Implementation}
\label{sec:implementation}

We implement \cascade{} on top of vLLM~\cite{vllm} and use LMCache~\cite{lmcache} to move KV blocks across HBM, DRAM and NVMe tiers. vLLM's continuous batching, chunked prefill, paged KV allocation and prefix-cache block hashing are reused unchanged. \cascade{} adds the TTFT estimator, the budget computation, the queue ordering of Section~\ref{sec:cd}, the budget-bounded prefetcher, and backfill admission. It needs no change to model weights and no extra hardware.
%
Besides vLLM's token sequence and block table, each request carries its class targets, $\hat T_r$, $B_r$, $a_r$, its prefix-hit profile, and the bounds $M_{r,k}$. $B_r(t)$ is derived each iteration (Eq.~\ref{eq:rem_budget_eq}) and not stored, so the only per-iteration write is the sort key.
%
The estimator and the ordering run on the CPU over queue occupancy and per-tier hit metadata. Each iteration makes one pass over the queue for $\rho$, one sort, and one linear walk for the backlog test (Algorithm~\ref{alg:cascade}, steps 2 and 3). This work overlaps with the GPU execution of the current batch, and its cost is negligible next to a decode step.
%
Reused prefix state stays in the deeper tiers until a request will consume it. A deep-tier hit is staged NVMe to DRAM to HBM, and \cascade{} allocates HBM only at the final DRAM-to-HBM hop, so HBM is not held for the whole multi-hop transfer. Blocks are reclaimed least recently used within each tier, which keeps hot prefixes in the fast tiers.
%
On top of vLLM's per-block hashing, a cluster-wide index records the tier that holds each cached block. A request's matched prefix length and per-block tiers form the hit profile that the estimator consumes (Section~\ref{sec:latency_estimator}) and that the prefetcher bounds with $M_{r,k}$ (Eq.~\ref{eq:max_restorable}).
%
Batch composition changes only at chunk boundaries, so \cascade{} never interrupts a running kernel. When the running requests outgrow HBM, the victim is chosen as in vLLM (the most recently admitted request). Its KV state is spilled to DRAM, and it re-enters the queue with the budget it has left and a fetch bound recomputed from that budget. The optional preemption of Section~\ref{sec:cd} acts at the same boundary.

\section{Evaluation}
\label{sec:evaluation}

\subsection{Methodology} \label{sec:methodology}

\paragraph{Models and Serving Cluster}

We evaluate \cascade across three representative LLMs: Qwen-2-72B~\cite{qwen}, Llama-3-70B, and Llama-3-405B~\cite{llama3}, served with a tensor parallelism (TP) degree of 4~\cite{tensorparallelism}. Model weights and KV cache blocks are maintained in FP16 precison. While Qwen-2-72B use employ Multi-Head Attention(MHA), Llama-3-70B and Llama-3-405B use Grouped-Query Attention (GQA)~\cite{gqa}. All hardware profiling is conducted on NVIDIA GB300 NVL72 nodes interconnected via Multi-Node NVLink (MNNVL)~\cite{gb200, mnnvl}. To evaluate large-scale production traces across multi-replica clusters, we extend the Vidur simulator~\cite{vidur} to model multi-tier memory hierarchies and LMCache~\cite{lmcache} for optimized KV cache data movement. We model the KV-data movement within simulator using the practical bandwidth achieved on physical GB300 NVL72 hardware across NVLink-C2C and MNNVL interconnects. To provide evaluation fidelity, we compare vidur simulated results against actual GPU runs across different models and context lengths in Sec~\ref{sec:eval_fidelity}.

\begin{scriptsize}
\begin{table}[t]
\centering
\caption{SLO targets}
\resizebox{1\columnwidth}{!}{
\begin{tabular}{ l | l | c c c c c | c }
\toprule
\multirow{2}{*}{\textbf{Model}} & \multirow{2}{*}{\textbf{Bucket}} & \multicolumn{5}{c|}{\textbf{TTFT (s)}} & \textbf{TPOT} \\
& & \textbf{ChatBot} & \textbf{API} & \textbf{Coder} & \textbf{Reasoning} & \textbf{Tool\&Agent} & \textbf{(ms)} \\
\midrule
\multicolumn{8}{c}{\textit{Flat-class SLOs}} \\
\midrule
Qwen-2-72B & Flat & 2.88 & 0.51 & 5.81 & 7.05 & 9.13  & 75  \\
Llama-3-70B  & Flat & 1.68 & 0.51 & 4.72 & 6.89 & 14.89  & 75 \\
Llama-3-405B & Flat & 9.12 & 1.82 & 18.33 & 24.94 & 37.64 & 250 \\
\midrule
\multicolumn{8}{c}{\textit{Request length-based SLOs}} \\
\midrule
\multirow{4}{*}{Qwen-2-72B}
  & p50      & 0.24 & 0.21 & 1.93  & 1.08 & 2.58 & \multirow{4}{*}{75} \\
  & p50--p75 & 1.08 & 0.48 & 5.64  & 2.24 & 4.56 & \\
  & p75--p90 & 3.16 & 0.80 & 7.61  & 7.37 & 10.38 & \\
  & p90+     & 8.82 & 2.49 & 9.28  & 17.47 & 69.65 & \\
\midrule
\multirow{4}{*}{Llama-3-70B}
  & p50      & 0.24  & 0.21 & 1.69  & 1.05  & 2.33 & \multirow{4}{*}{75} \\
  & p50--p75 & 0.97  & 0.46 & 5.44  & 2.19  & 4.35 &  \\
  & p75--p90 & 2.90  & 0.80 & 7.32 & 7.36 & 9.80 & \\
  & p90+     & 7.43 & 2.46 & 8.92 & 17.22 & 58.48 & \\
\midrule
\multirow{4}{*}{Llama-3-405B}
  & p50      & 0.83  & 0.62 & 6.68  & 3.80  & 9.06 & \multirow{4}{*}{250} \\
  & p50--p75 & 3.80  & 1.62 & 19.62 & 8.01  & 15.45 & \\
  & p75--p90 & 11.15  & 2.89 & 26.31 & 26.13 & 36.15 & \\
  & p90+     & 29.95 & 8.94 & 31.99 & 60.03 & 222.36 & \\
\bottomrule
\end{tabular}}
\label{tab:slo_target}
\end{table}
\end{scriptsize}

\paragraph{Workloads and Traces}
\label{sec:workloads}

\begin{figure*}[!t]
  \centering
  \includegraphics[width=1\textwidth]{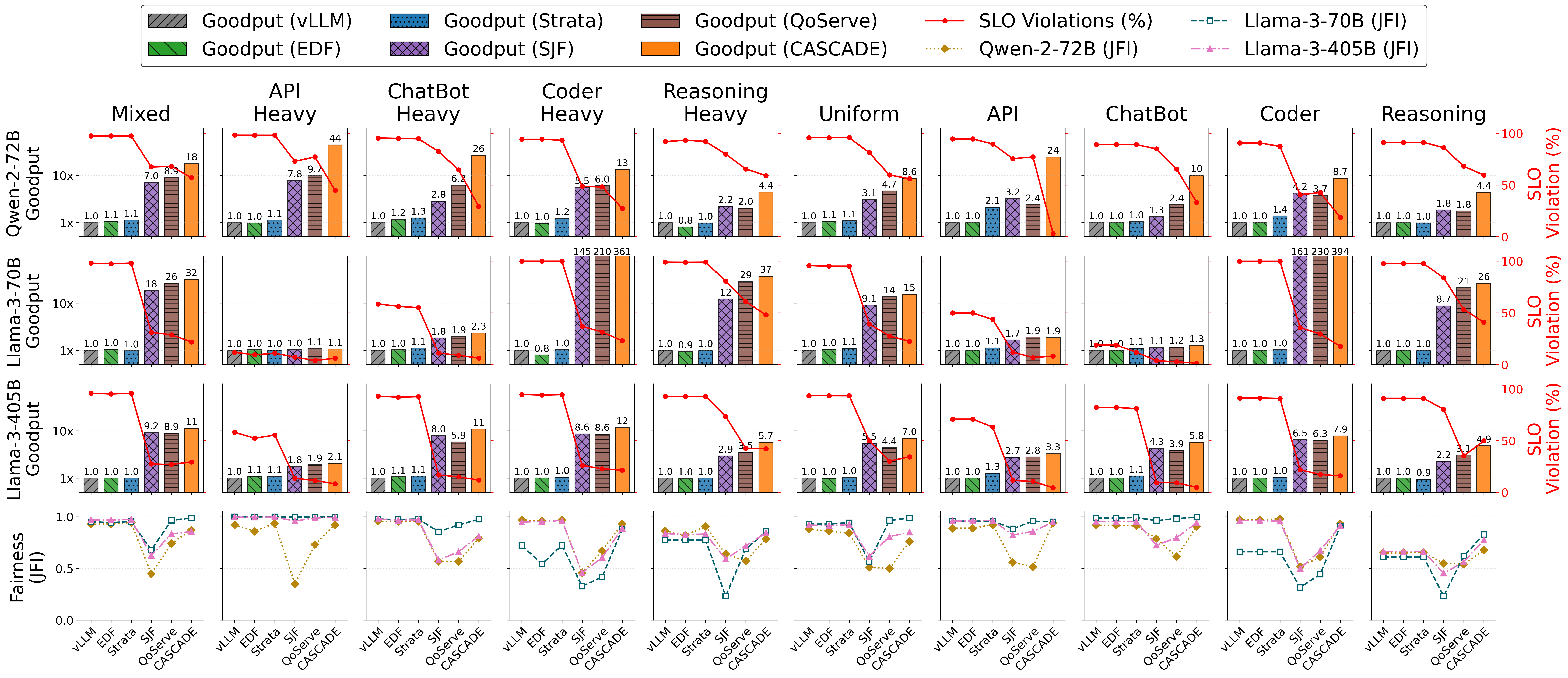}
   \caption{Serving performance across three models and ten traces under \emph{flat-class SLOs}.
  Bars show goodput (tokens/s) normalized to \textsc{vLLM}; markers show SLO violation
  rate (right axis). On average, \cascade improves goodput by \avggoodputimprvllmflatall over
  \textsc{vLLM} and \avggoodputimprqoserveflatall over QoServe, and reduces SLO violations by
  \avgsloviolationredvllmflatall and \avgsloviolationredqoserveflatall, respectively, while
  remaining fair across all requests.}
   \label{fig:goodput_vs_slo}
\end{figure*}

To evaluate \cascade under realistic multi-tier KV caching dynamics, we utilize open-sourced production traces collected from a Qwen model serving cluster on Aliyun Bailian~\cite{traces}. The dataset consists of a two-hour sampled trace containing real request arrival timestamps and exact prefix block hash matches, capturing four dominant LLM application classes: \emph{ChatBot} (interactive conversation), \emph{API} (api-driven task automation), \emph{Coder} (code generation), and \emph{Reasoning} (multi-step thinking workloads). Using these base application traces, we construct five workload mixtures to evaluate diverse deployment scenarios: (1) \emph{Mixed}, preserving the authentic production distribution across all four classes; (2) four \emph{App-Heavy} workloads, where a target application accounts for 50\% of total traffic and the remaining 50\% is split evenly among the other three; and (3) \emph{Uniform}, containing an equal 25\% split per application class. To evaluate agentic workloads~\cite{agenticaiworkloadcharacteristics, generative_agents, agenticcostdynamicreasoningdemystifying}, we employ MoonCake's~\cite{mooncake} \emph{Tool\&Agent} (tool \& agents workload) one-hour trace for agentic workloads evaluation.  Table~\ref{tab:traces} summarizes the key characteristics of each trace, including input/output token length distributions, total request counts, and multi-turn ratios.

\paragraph{Service Level Objective (SLO) Targets}

To provide rigorous quality-of-service assurances, serving systems must satisfy strict latency targets defined by SLAs. Both, flat-class~\cite{mooncake, qoserve} and tiered~\cite{birds, aladdin, polyserve, sloawarescheduling, binaryprioritiesmultitiersla} (request's input length proportional) SLA's have been adopted in production serving. For flat-class SLO's, following prior work~\cite{mooncake}, we define the 90th-percentile Time-to-First-Token ($\text{TTFT}_{p90}$) and Time-Per-Output-Token ($\text{TPOT}_{p90}$) flat-class SLOs relative to isolated baseline execution delays. Specifically, we set the normalized targets to $\text{TTFT}_{p90} = 10\times$ and $\text{TPOT}_{p90} = 5\times$, meaning $90\%$ of requests in a given class must achieve a TTFT and TPOT within $10\times$ and $5\times$ of a single request executing on identical hardware without batching or cross-tenant interference. 
While for tiered SLO, we set TTFT SLOs by prefill-length. The TTFT SLO of a bucket is $10\times$ the $p90$ prefill time of the requests in that bucket.
Table~\ref{tab:slo_target} summarizes the flat and request length-based SLO-targets  targets for each model architecture, workload trace, and prefill-length bucket.

\paragraph{Baselines}
We implement \textsc{Cascade} on top of the \textsc{vLLM}-v1 inference engine~\cite{vllm} and evaluate it state-of-the-art scheduling policies: 
(1)~\textbf{\textsc{vLLM}(FCFS)}, the default scheduling policy for \textsc{vLLM}~\cite{vllm}, SGLang\cite{sglang} and TensorRT-LLM~\cite{trtllm}, which dispatches requests across all workload classes in First-Come, First-Served order; 
(2)~\textbf{EDF}, an Earliest Deadline First policy that introduces deadline awareness by prioritizing requests based on their target TTFT SLO deadlines; 
(3)~\textbf{SJF}, a Shortest Job First policy that prioritizes requests with shorter input sequence lengths to minimize prefill compute overhead;
(4)~\textbf{Strata}~\cite{strata}, a KV-aware scheduler that mitigates delay hits while balancing batches to hide cache-loading latency; and
(5)~\textbf{QoServe}~\cite{qoserve}, an SLO-aware scheduler that employs hybrid prioritization and an eager relegation policy for graceful service degradation under overload.
For a fair comparison, all policies and \textsc{Cascade} are evaluated under identical cluster settings within our extended Vidur simulator framework. All configurations use the same \emph{global scheduler}, which routes requests based on HBM prefix cache hits~\cite{preble} and GPU prefill load, and enforces prefill-decode (PD) co-location with chunked prefills~\cite{sarathi} (chunk size of 512 tokens, maximum batch size of 128 - QoServe employs dynamic chunk size) to exploit slack within TPOT SLO. The LMCache~\cite{lmcache} chunk size is set to 512 as default across all experiments. Unless otherwise stated, our evaluation on the \emph{Mixed} workload trace deploys a cluster capacity comprising 9 instances of Qwen-2-72B, 4 instances of Llama-3-70B, and 12 instances of Llama-3-405B with keeping overall serving system under high load state across all workloads.
We target the standard deployment model of vLLM~\cite{vllm}, where an instance (TP=4 on GB300, roughly 1~TB of HBM) hosts one full model instance and serves each request end to end. Instances are provisioned so that the longest supported context fits.

\subsection{Serving Efficiency under Production Traces}
\label{sec:eval_efficiency}
We first evaluate \textsc{Cascade} against the baselines using the arrival timestamps from the production traces, then decompose the result to show which serving overheads it removes.

\subsubsection{LLM Serving under flat-class SLOs}
\label{sec:eval_goodput}

Relying on a single LLM serving metric while serving LLM requests in production setting while complying with class-based SLOs would be misleading; goodput, SLO violations, fairness, and latency must be read together.
Figure~\ref{fig:goodput_vs_slo} reports goodput, SLO violations, fairness, and median TTFT for three models across ten production traces under \emph{flat-class SLOs}.
Goodput and violation counts alone are insufficient because requests within a class can differ in input length by orders of magnitude: a violation on a 1k-token request is not equivalent to one on a 128k-token request, whose execution time is far longer.
We therefore also report fairness, measured by Jain's Fairness Index over per-class SLO attainment; a value of one means all classes attain their SLOs at the same rate.

\begin{figure}[t]
  \centering
  \includegraphics[width=1\columnwidth]{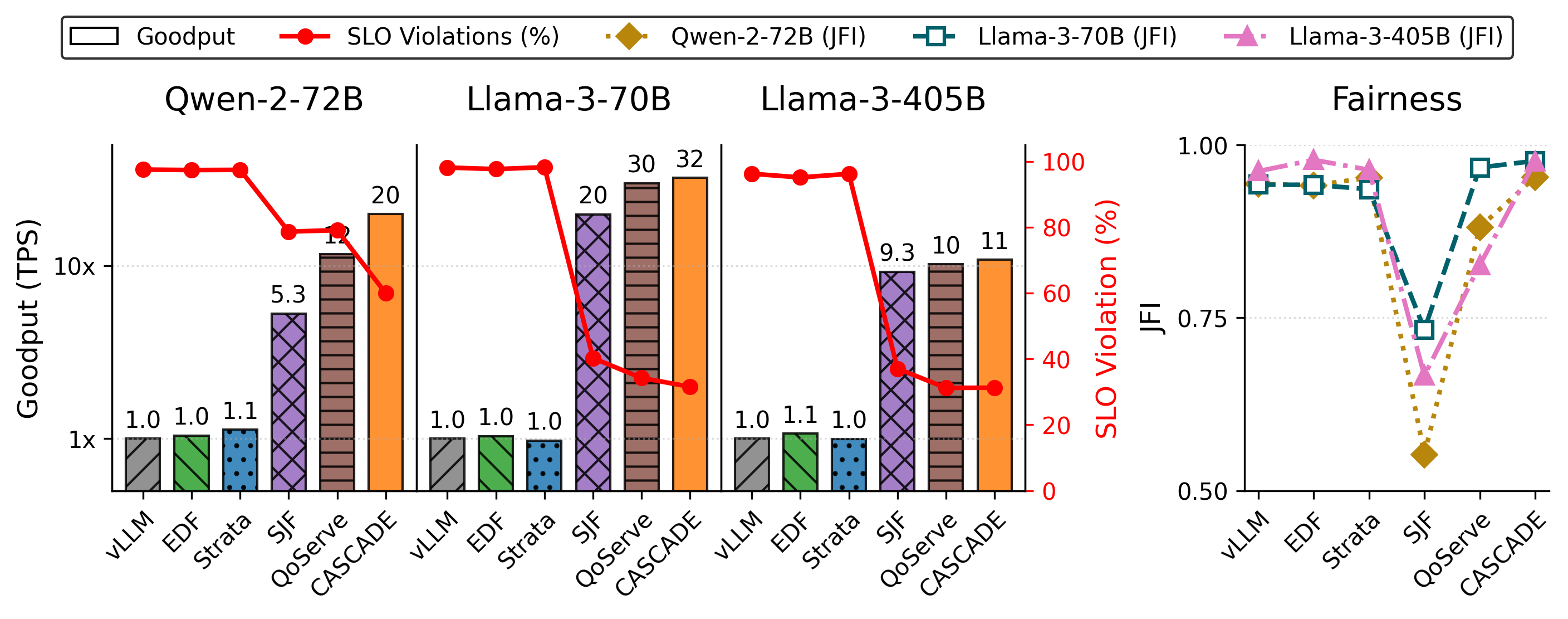}
  \caption{LLM serving metrics for \emph{Mixed trace} with \emph{request length-based SLOs}. Goodput normalized to \textsc{vLLM}.}
\label{fig:goodput_vs_slo_percentile}
\end{figure}

\begin{figure}[t]
  \centering
  \includegraphics[width=1\columnwidth]{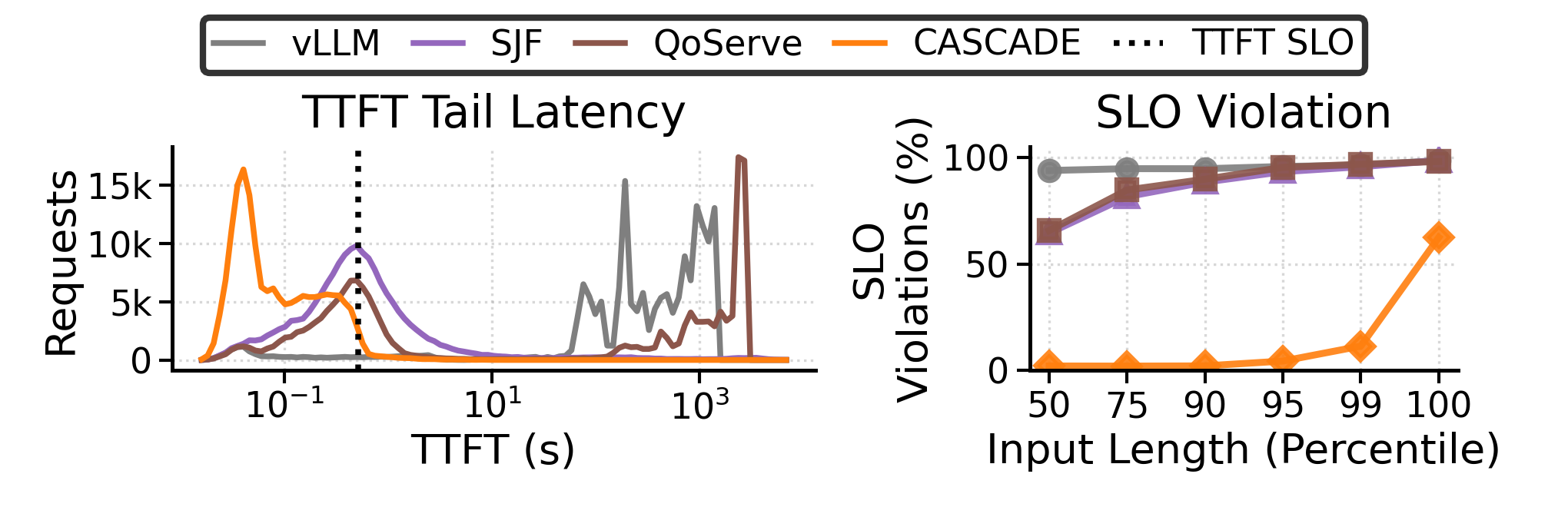}
  \vspace{-0.5cm}
    \caption{TTFT tail latency and SLO violation rate grouped by input-sequence-length percentile for the
    \emph{API} trace with Qwen-2-72B model.}
\label{fig:prompt_slo_violation}
\end{figure}

On average, \textsc{Cascade} improves goodput by \avggoodputimprvllmflatall over \textsc{vLLM} and reduces violations by \avgsloviolationredvllmflatall. Against the state-of-the-art SLO-aware QoServe, it improves goodput by \avggoodputimprqoserveflatall and reduces violations by \avgsloviolationredqoserveflatall.
Strata, the recent KV-aware baseline, offers little benefit over \textsc{vLLM} because it retains FCFS scheduling. SJF achieves strong goodput and violation numbers but is unfair on all workloads, as it prioritizes short requests over long ones.
On \emph{Coder} and \emph{Coder Heavy}, the Llama-3-70B baselines using FCFS, EDF, and Strata perform worst under high load, which inflates the relative goodput of SJF, QoServe, and \textsc{Cascade}; we treat these as outliers.
For Llama-3-405B on \emph{Uniform} and \emph{Reasoning}, \textsc{Cascade} has slightly more violations than QoServe but higher goodput. This is explained by fairness: QoServe meets more SLOs by deprioritizing long-context requests in favor of short ones, whereas \textsc{Cascade} treats all length classes fairly and achieves better goodput and fairness overall, reinforcing that no single metric suffices.

Per model, \textsc{Cascade} improves goodput over QoServe by \avggoodputimprqoserveflatqwen, \avggoodputimprqoserveflatllamasmall, and \avggoodputimprqoserveflatllamabig for Qwen-2-70B, Llama-3-70B, and Llama-3-405B, respectively. The gain is largest on Qwen, whose MHA design yields a larger KV cache and thus more data-movement time. We expect the benefit to grow for larger models with even larger KV caches.

\subsubsection{LLM Serving under request-length based SLOs}
\label{sec:percentile_slo}

Figure~\ref{fig:goodput_vs_slo_percentile} reports goodput, SLO violation rate, and fairness under \emph{request-length-based SLOs} on the \emph{Mixed} workload.
Here, SLOs are tighter for shorter requests and looser for longer ones in comparison to \emph{flat-class SLOs}. \textsc{Cascade} exploits this headroom by scheduling short requests within the latency budget of long requests, and outperforms all baselines.
Compared to QoServe, \textsc{Cascade} improves goodput by \avggoodputimprqoservepercentileall and reduces SLO violations by \avgsloviolationredqoservempercentileall on average, showing that it adapts to both \emph{flat} and \emph{request-length-based} SLOs all while providing fairness.

\subsubsection{Agentic Workloads - Transient Overload}

\begin{figure}[t]
  \centering
  \includegraphics[scale=1]{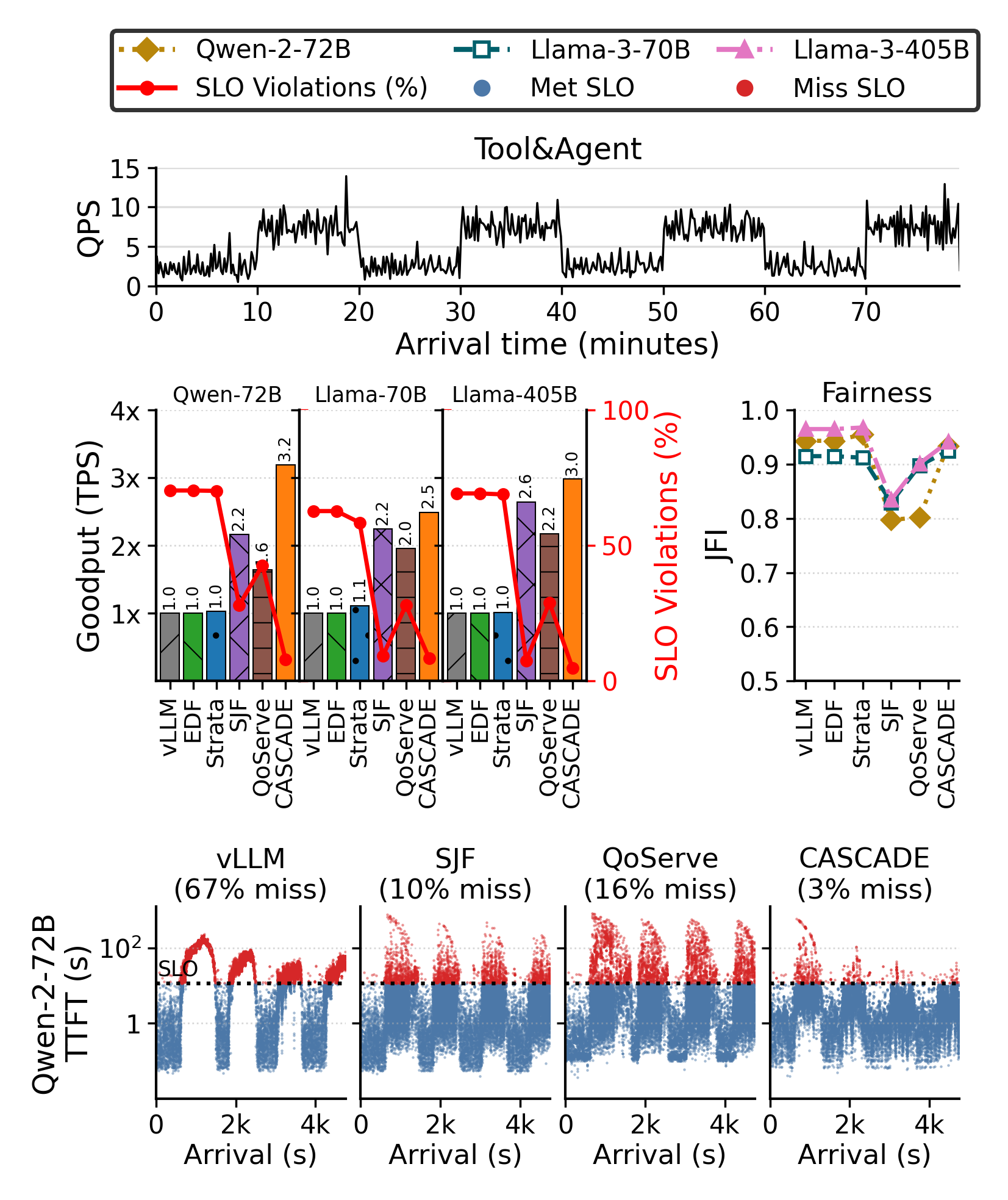}
    \caption{Agentic workload with \emph{Tool\&Agent} trace. Per-request TTFT versus arrival rate and SLO compliance.}
\label{fig:toolagent}
\end{figure}

Agentic workloads, composed of agents and subagents invoking tools, cause transient overloads: bursts of requests arrive together, followed by idle periods while agents await the next user step.
To study \textsc{Cascade} under such conditions, we use \emph{Tool\&Agent} trace, which contains transient overloads and long requests with high turn ratios and cross-session prefix hits.
Figure~\ref{fig:toolagent} shows the request arrival pattern together with goodput, SLO violations, fairness, and per-request TTFT.
\textsc{Cascade} consistently outperforms all baselines by exploiting KV-awareness and SLO-aware latency budgets, while remaining fair across request lengths. In contrast, SJF and QoServe prioritize short requests during overload bursts, which leads to unfairness.

\subsubsection{Sources of SLO Violation} \label{sec:eval_breakdown}

\begin{figure}[t]
  \centering
  \includegraphics[scale=1]{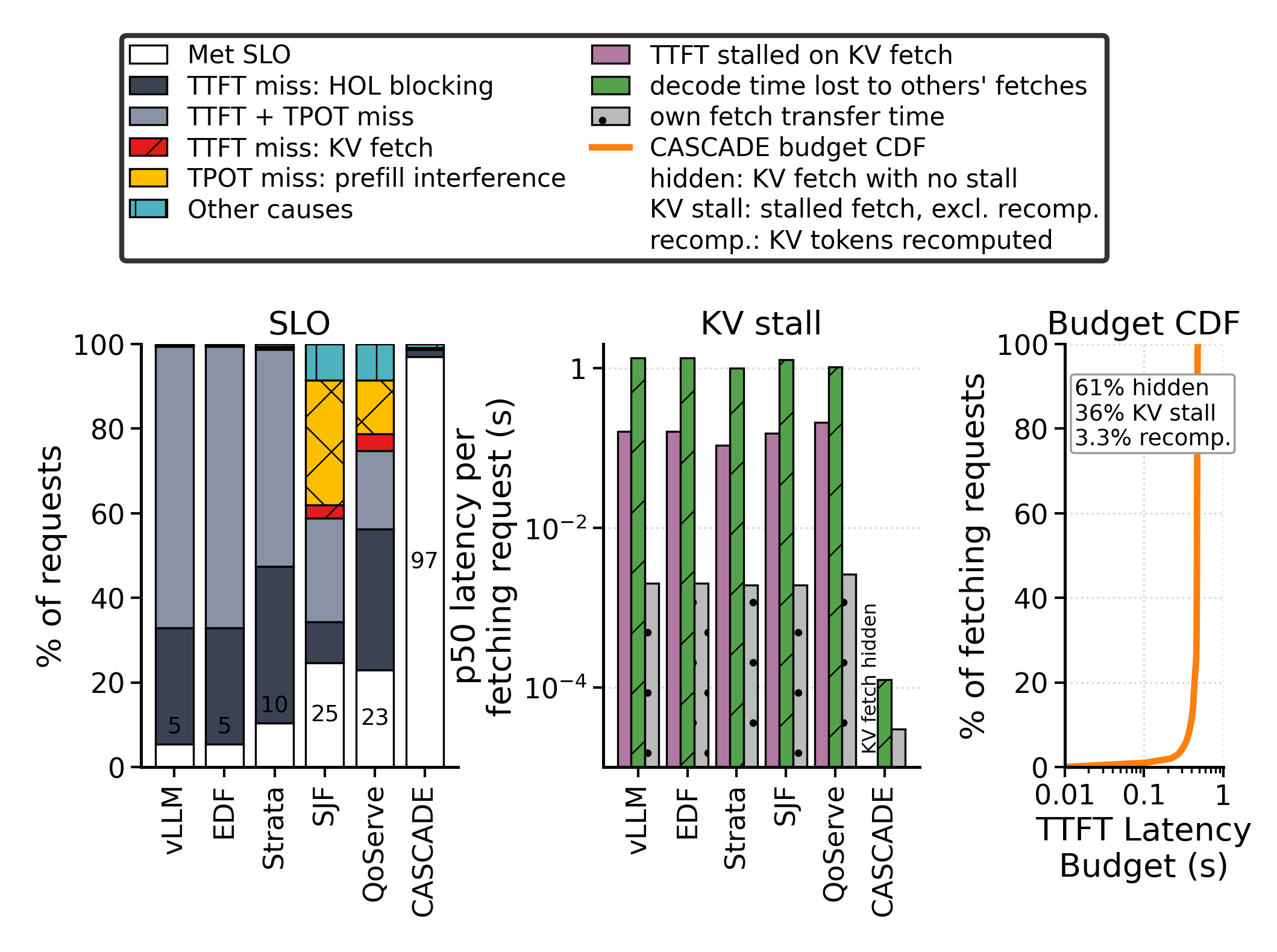}
    \caption{SLO violations and KV stall breakdown over \emph{API} trace for Qwen-2-72B model.}
\label{fig:slo_violations_breakdown}
\end{figure}

To explain why \textsc{Cascade} lowers SLO violations and improves goodput, Figure~\ref{fig:slo_violations_breakdown} breaks down SLO violations by cause, reports the median serving stall due to KV fetch, and shows the fraction of requests whose KV fetch \textsc{Cascade} hides within its latency budget.
Most violating requests miss both TTFT and TPOT SLOs. For TTFT, head-of-line (HOL) blocking is the largest cause, followed by KV fetch. For TPOT, the main cause is prefill interference, which mostly stems from KV fetch.
Because of PD colocation, a KV fetch stalls not only effects the request being scheduled but also the other prefill and decode requests in the same batch.
\textsc{Cascade} eliminates both causes. It hides KV fetch latency while a request waits in the scheduling queue, and it removes HOL blocking by scheduling shorter requests when a longer request blocks the queue. Only tier-2 requests (those with negative budget) still face HOL blocking.
\textsc{Cascade} hides KV fetch for 61\% of requests. The remaining 39\% incur a small stall, and of these, 3.3\% are recomputed because fetching from lower-tier memory would take longer under the current load.
As a result, the p90 KV-fetch stall in TTFT is only 110\,ms.

\subsubsection{Tail Latency and SLO Violations by Request Size} \label{sec:eval_tail}

Figure~\ref{fig:prompt_slo_violation} shows the TTFT distribution and SLO violation rate for request groups binned by input length, for Qwen-2-72B on the \emph{API} trace.
\textsc{vLLM} serves requests in arrival order, so most requests suffer HOL blocking, producing peaks at high TTFT. SJF prioritizes short requests, creating a peak around the SLO with its right half in violation. QoServe eagerly relegates long requests, producing a large peak at very high TTFT and a smaller one around the SLO.
\textsc{Cascade} exploits per-request latency budgets and reduces KV stalls, which keep overall TTFT within SLO. On some workloads it relegates requests with negative budgets, producing a small peak at high TTFT; otherwise its mass stays within the SLO.
Consequently, \textsc{vLLM} violates SLOs regardless of request length. SJF and QoServe violate few short requests but nearly all long ones. \textsc{Cascade} violates only negative-budget requests, which tend to be very long, and treats all other requests fairly regardless of length.

\begin{figure}[t]
  \centering
  \includegraphics[scale=0.9]{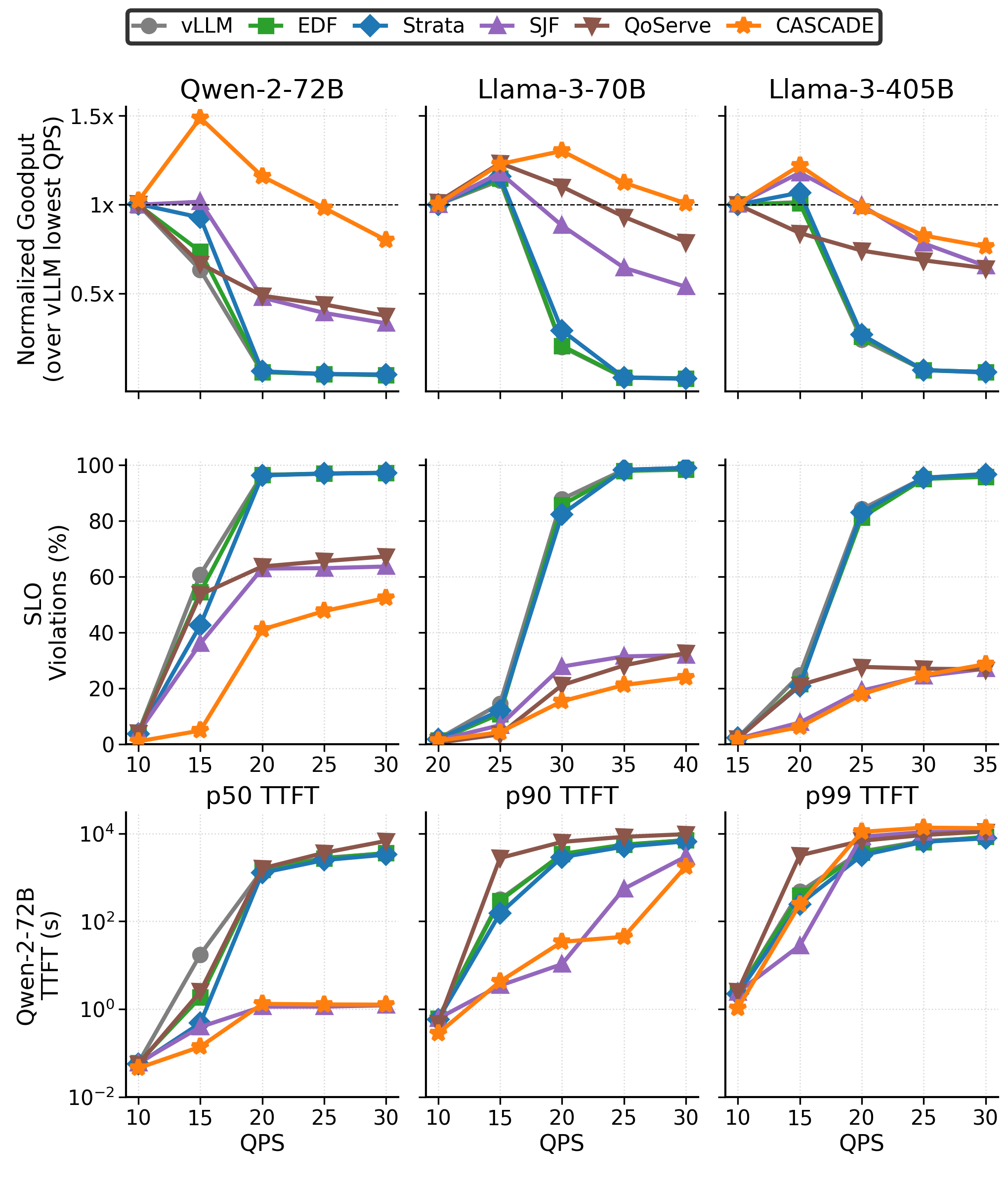}
    \caption{Goodput, SLO violation and TTFT under varying load with \emph{Mixed} trace.}
\label{fig:load_goodput}
\end{figure}

\subsubsection{Sensitivity to Offered Load} \label{sec:eval_load}

Figure~\ref{fig:load_goodput} shows goodput, SLO violation rate, and TTFT as we increase load on the serving cluster. The cluster is heavily loaded at 20, 30, and 25 QPS for Qwen-2-72B, Llama-3-70B, and Llama-3-405B, respectively.
At high QPS, \textsc{vLLM}, EDF, and Strata all exceed 80\% SLO violations, while SJF and QoServe performs comparatively better. \textsc{Cascade} adapts to the load and keeps median TTFT consistent as QPS increases.
Higher load also increases traffic on the interconnects used to fetch prefix KV, so a smaller fraction of KV fetches can be hidden and more are exposed or fall back to recomputation.

\subsection{Ablation Studies}

\subsubsection{Sensitivity to CPU-GPU link} \label{sec:eval_link}

To study the effect of the CPU-GPU link on \textsc{Cascade}, we sweep it from PCIe5 (128 GB/s) to NVLink-C2C (1800 GB/s), assuming all request KV is already offloaded to CPU DRAM.
\textsc{Cascade} prefetches a request's prefix while the request queues, guided by its latency budget, so a fetch is hidden if it completes before admission.
On PCIe5, 58\% of requests reusing a DRAM prefix hide the fetch, and the remaining 42\% stall the batch. NVLink-C2C raises the hidden share to 74\% and, more importantly, shortens the remaining stalls: p50 falls from 0.40\,s to 0.03\,s, p90 from 1.62\,s to 0.12\,s, and p99 from 4.56\,s to 0.36\,s, a roughly 13$\times$ reduction at each percentile.

\begin{figure}[t]
  \centering
  \includegraphics[scale=0.9]{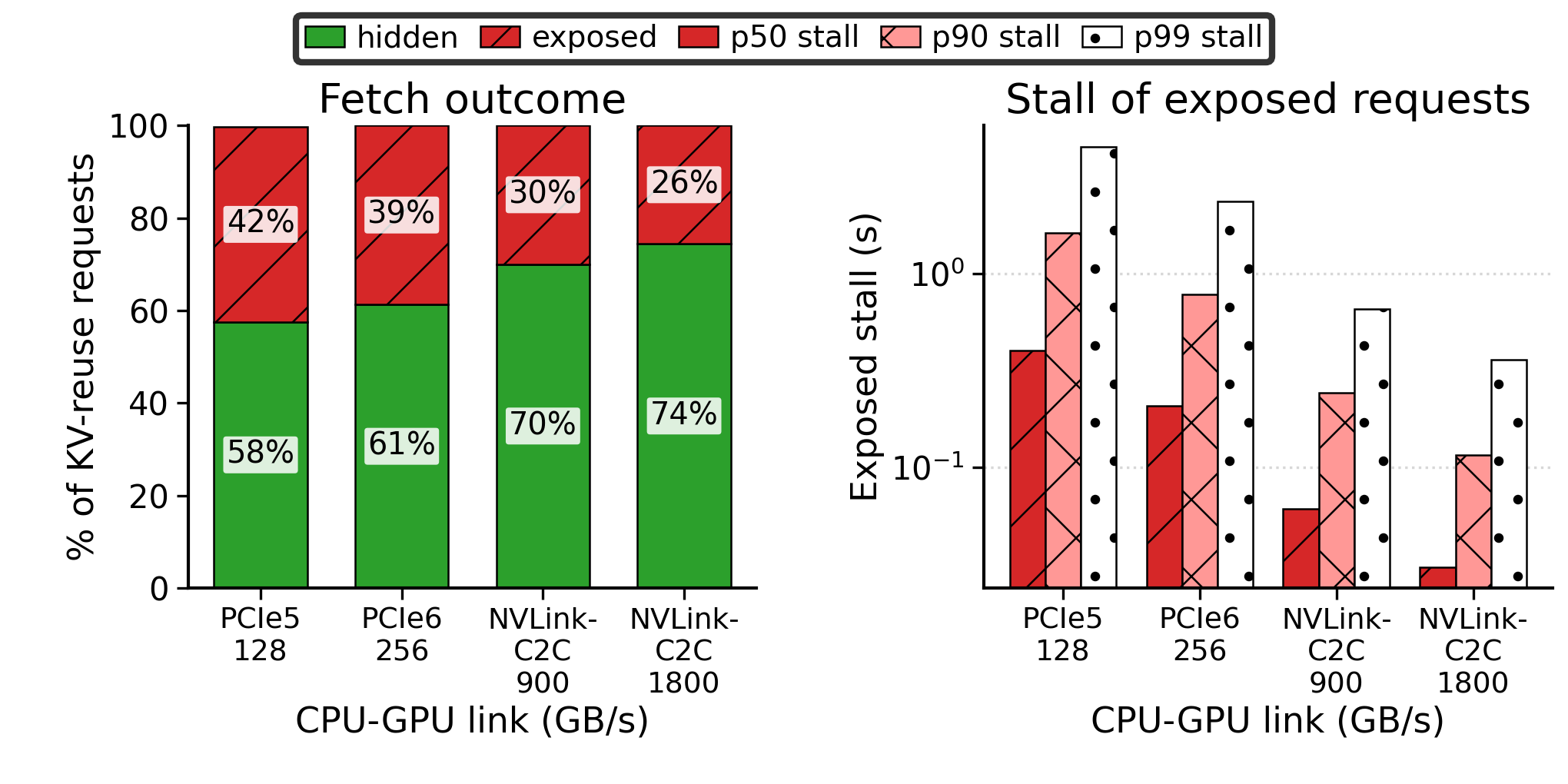}
    \caption{Effect of the CPU-GPU link on KV prefetch in \cascade (Qwen-2-72B, Tool\&Agent trace). \textbf{Left:} share of requests that reuse a DRAM prefix. \textbf{Right:} p50, p90 and p99 admission stall of the exposed requests.}
\label{fig:link_ablation}
\end{figure}

\begin{table}[t]
\begin{minipage}[b]{0.56\columnwidth}
\caption{\cascade{} ablation on Llama-3-70B, \emph{Mixed} trace. Configs are cumulative.}
\vspace{-0.3cm}
\label{tab:ablation}
\end{minipage}\hfill
\begin{minipage}[b]{0.40\columnwidth}
\caption{Evaluation fidelity on GB300 (TP4), 2k to 128k tokens.}
\vspace{-0.3cm}
\label{tab:fidelity}
\end{minipage}

\vspace{0.5ex}
\centering
\setlength{\tabcolsep}{3pt}
\resizebox{\columnwidth}{!}{%
\begin{tabular}[b]{@{}lcccc@{}}
\toprule
\multirow{2}{*}{Config} & Goodput & Viol. & \multicolumn{2}{c}{TTFT (s)} \\
\cmidrule(l){4-5}
 & (k tok/s) & (\%) & p50 & p90 \\
\midrule
Backfill         & 43.6 & 41.1 & 1.13 & 4{,}977 \\
+ Cost-aware     & 49.0 & 24.3 & 0.52 & 3{,}669 \\
+ Backlog-aware  & 50.8 & 24.0 & 0.52 & 3{,}587 \\
\bottomrule
\end{tabular}%
\hspace{1.5em}%
\begin{tabular}[b]{@{}lcc@{}}
\toprule
\multirow{2}{*}{Model} & \multicolumn{2}{c}{Mean abs. error (\%)} \\
\cmidrule(l){2-3}
 & Prefill & Decode \\
\midrule
Qwen-2-72B     & 5.2 & 15.2 \\
Llama-3-70B    & 4.3 & 18.7 \\
Llama-3.1-405B & 1.2 & 9.1 \\
\bottomrule
\end{tabular}}
\end{table}

\subsubsection{\cascade Techniques} \label{sec: cascade_ablation}

We ablate \textsc{Cascade} on Llama-3-70B with \emph{Mixed} trace, adding backfill, cost-aware priority, and backlog-aware budget in turn (Table~\ref{tab:ablation}).
The gains come from the two ordering mechanisms. Cost-aware priority raises goodput by 12.4\% and cuts SLO violations from 41.1\% to 24.3\%, while halving median TTFT (1.13\,s to 0.52\,s) and lowering p90 TTFT by 26\%.
Backlog slack adds a smaller gain: goodput rises another 3.6\% and violations fall to 24.0\%. In total, the full system (a3) achieves 16.5\% higher goodput than a1 with 17.1 percentage points fewer SLO violations.

\subsubsection{Evaluation Fidelity} \label{sec:eval_fidelity}

To validate simulation fidelity, we compare Vidur~\cite{vidur} against real GB300~\cite{mnnvl} node on requests with prompts of 2k to 128k tokens and report median wall-clock time over three repetitions. Table~\ref{tab:fidelity} lists the errors.
\emph{Prefill.} The mean absolute error (MAE) over seven prompt lengths is 1.2\% for Llama-3-405B, 4.3\% for Llama-3-70B, and 5.2\% for Qwen-2-72B.
\emph{Decode.} The MAE over the same prompt lengths is 9.1\% for Llama-3-405B, 15.2\% for Qwen-2-72B, and 18.7\% for Llama-3-70B.
\section{Related Work}
\label{sec:related_work}

Service-Level Objectives (SLOs) have long served as a foundational abstraction for resource allocation and task scheduling across diverse computing domains, including mobile application runtimes~\cite{timecard}, data-parallel processing clusters~\cite{jockey}, and latency-sensitive datacenter networks~\cite{betterneverthanlate}. Applying SLO-driven scheduling to LLM serving, however, introduces distinct challenges due PD-colocation  challenges. Recent works have explored this design space along several orthogonal dimensions. AlpaServe~\cite{alphaserve}, Nitsum~\cite{nitsum} optimizes model-parallelism strategies to satisfy heterogeneous SLO requirements, while Learned Best-Effort LLM Serving~\cite{learnedbesteffortllmserving} applies reinforcement learning to route requests across model variants to balance accuracy against latency targets. 

Another line of research explores chunked execution and opportunistic batching to balance prefill throughput against decode latency bounds. Building on the chunked prefill model of Sarathi~\cite{sarathi}, Medha~\cite{medha}, PolyServe~\cite{polyserve}, and QoServe~\cite{qoserve} adaptively scale chunk granularities to prevent accumulating attention overhead from breaching Time-Between-Tokens ($\text{TBT}$) limits. QoServe leverages request-level latency slack for chunk size tuning. Conserve~\cite{conserve} and SageServe~\cite{sageserve} employ reactive heuristics—such as scheduling background tasks when interactive load drops below fixed thresholds (e.g., $60\%$) or applying reactive preemption during load spikes. However, these systems implicitly assume that incoming prompts require full GPU recomputation, treating $\text{TTFT}$ purely as a compute-bound phase while ignoring deep-tier KV restoration overheads.
SLO-aware schedulers order and size work by latency targets but treat a
request's cost as its prompt or output length. Sarathi-Serve~\cite{sarathi}
bounds decode stalls with chunked prefill but is SLO-agnostic.
SLOs-Serve~\cite{sloawarescheduling} admits only the request set whose multi-stage SLOs are attainable, and JITServe~\cite{jitserve} allocates just enough serving bandwidth to meet each SLO from imprecise request information. None of them models where a request's prefix KV resides, so a prompt cached in HBM and one cached on NVMe are priced alike. KV-aware systems take the opposite trade-off.
Strata~\cite{strata} hides hierarchical KV loading behind compute with
delay-hit deferral and balanced batches, but serves requests in FIFO order without SLOs. Mooncake~\cite{mooncake} selects prefill instances by cache hits
and rejects requests early from predicted load in a disaggregated cluster. Llumnix~\cite{llumnix} migrates requests between instances using memory usage and binary priorities. Simple is Better~\cite{lmetric} routes by
multiplying a KV-hit indicator with batch size. \cascade bridges both lines within a single engine: it derives a per-request latency budget from a prefix-cache-tier-aware TTFT estimate, and restores prefix KV from deeper-tier only of budget allows.

VTC~\cite{vtc} defines fairness across clients: it counts the input and output
tokens served to each client and bounds the service difference between
backlogged clients, without latency targets. \cascade targets a different axis,
fairness across request lengths. The two are orthogonal: a per-client token counter such as VTC could be layered on top of \cascade.

\section{Conclusions}

Serving a language model request under a latency SLO involves more than running the
model: the request also incurs queueing delay, batching interference, and the cost of moving reused KV cache into GPU memory, and all of it is charged against the same SLO.
The time an SLO leaves for this overhead is the request's latency budget, which
differs by orders of magnitude across requests. We present \textsc{Cascade}, a serving system that estimates each request's latency budget at runtime and treats it as a single quantity that governs two decisions handled in isolation today: how requests are scheduled, and how their reused KV cache is placed across the memory hierarchy.
By steering overhead toward requests with budget to spare, \textsc{Cascade} improves
SLO attainment and throughput together while being fair, across a range of models, applications, and
request sizes, and without additional hardware.

\balance
\bibliographystyle{IEEEtranS}
\bibliography{refs}

\end{document}